\documentclass[fontsize=11pt,a4paper]{arxiv-article}
\bibliography{References.bib,not-arxiv.bib}

\usepackage{Definitions}

\usepackage{hyperref}
\graphicspath{{./plots/}}

\allowdisplaybreaks[1]
\newcommand{\vect}[1]{\mathbf{#1}}
\newcommand{\proptime}{\tau}
\newcommand{\Kdiag}{K}
\newcommand{\Koff}{\tilde{K}}
\newcommand{\bbdiag}{\mathbb{B}}
\newcommand{\bboff}{\tilde{\mathbb{B}}}

\usepackage{cancel}

\newcommand{\OfficialTitle}{
	The unitary Fermi gas\\at large charge and large N
}
\title{\setstretch{1.4}
	{\color{Thoughtless}\textls[-20]{\OfficialTitle}}
}

\hypersetup{pdfauthor={Simeon Hellerman, Daniil Krichevskiy, Domenico Orlando, Vito Pellizzani, Susanne Reffert, Ian Swanson},pdftitle={\OfficialTitle},%
	colorlinks=true,linkcolor=ThoughtYouWere,citecolor=ThoughtYouWere,urlcolor=ThoughtYouWere}

\author{%
	\begin{minipage}{.94\textwidth}
		\begin{center} \dosserif%
			{\small
				\textbf{Simeon Hellerman}\textsuperscript{\ding{74}},
				\textbf{Daniil Krichevskiy}\textsuperscript{\ding{88}\ding{73}},
				\textbf{Domenico Orlando}\textsuperscript{\ding{72}\ding{73}},
				\textbf{Vito Pellizzani}\textsuperscript{\ding{73}},
  				\textbf{Susanne Reffert}\textsuperscript{\ding{73}} and
  				\textbf{Ian Swanson}
			}
		\end{center}
  		\authorBlock{\ding{74}}{\dosserif{}%
			Kavli Institute for the Physics and Mathematics of the Universe (WPI)\\ The University of Tokyo\\ Kashiwa, Chiba 277--8582, Japan}
        \authorBlock{\ding{88}}{\dosserif{}%
			Faculty of Science and Technology\\ University of Stavanger,\\ Kjell Arholms gate 41, 4021 Stavanger, Norway}
		\authorBlock{\ding{73}}{\dosserif{}%
			Albert Einstein Center for Fundamental Physics\\
			Institute for Theoretical Physics, University of Bern,\\
			Sidlerstrasse 5, CH-3012 Bern, Switzerland}
			\authorBlock{\ding{72}}{\dosserif{}%
			INFN sezione di Torino.\\
			via Pietro Giuria 1, 10125 Torino, Italy}
	\end{minipage}
}

\date{}

\begin{document}

\numberwithin{equation}{section}

\begin{titlepage}

	\maketitle

	\thispagestyle{empty}

	\vfill\dosserif{}

	\abstract{%
		\normalfont{}\noindent{}%
		We study the unitary Fermi gas in a harmonic trapping potential starting from a microscopic theory in the limit of large charge and large number of fermion flavors $N$.
      In this regime, we present an algorithmic procedure for extracting data from perturbation theory, order-by-order, without the need for other assumptions.
      We perform a gradient expansion in the interior of the particle cloud, sufficiently far from the cloud edge where the particle density drops rapidly to zero.
      In this latter region we present the first microscopic computation characterizing the contribution of the edge terms.
      The microscopic theory reproduces the predictions of the superfluid \acs{eft}, including the action, the form of the gap equation, and the energy of the system in a harmonic trap (which maps, via the non-relativistic state-operator correspondence, to the scaling dimension of the lowest operator of charge $Q$).
      We additionally give the Wilsonian coefficients at leading order in N up to \acs{nnlo} in the large-charge expansion. 
	}

\end{titlepage}

\setstretch{1.1}
\tableofcontents

\newpage

\section{Introduction}%
\label{sec:Introduction}

The large-charge expansion has been shown to lead to important simplifications in strongly-coupled relativistic \acp{cft}~\cite{Hellerman:2015nra,Monin:2016jmo,Alvarez-Gaume:2016vff,Gaume:2020bmp}.
In cases with a unique vacuum at large charge, the physics is that of a superfluid, and an \ac{eft} can be formulated as an expansion in inverse powers of the charge.
This \ac{eft} can be used for example to compute the energy of the ground state on the cylinder $\mathbb{R}\times S^d$ in a sector of fixed charge \(Q\), which gives, via the state-operator correspondence, the scaling dimension of the lowest operator of charge \(Q\).

\medskip
The large-charge expansion also renders useful simplifications for \ac{nrcft}~\cite{Favrod:2018xov,Kravec:2018qnu}. Here, the underlying symmetry is not the conformal algebra but the Schrödinger algebra (summarized, \emph{e.g.}, in~\cite{Nishida:2010tm,Favrod:2018xov,Kravec:2018qnu}), which naturally contains a conserved $U(1)$ charge (namely, particle number).
As in the relativistic case, the symmetry completely determines the form of the two-point function up to the scaling dimension.
Moreover, the notion of a state-operator correspondence persists for \ac{nrcft}: The scaling dimension of a given operator is the same as the energy of the corresponding state for the system in a harmonic trapping potential~\cite{Nishida:2010tm}.

\medskip
The unitary Fermi gas is an interesting and experimentally accessible example of 
a system with Schrödinger symmetry.
It can be realized in the laboratory via an ultracold Fermi gas in an optical trap~\cite{Strinati:2018wdg,PhysRevLett.92.120403}. By tuning an external magnetic field to a Feshbach resonance, the scattering length becomes infinite at the so-called unitary point, where the system acquires Schrödinger symmetry. 
 Ultra-cold Fermi gases in the presence of a {broad} Feshbach resonance can be described by a contact interaction. One requires that the gas is sufficiently dilute, which is the typical case in experiments~\cite{Giorgini:2008zz}. 
 This amounts to the inter-particle distance being much larger than the range of the inter-atomic potential, so that the details of the interactions are not important, and it is safe to focus on s-wave scattering.%
\footnote{Some nuclear systems feature a BCS--BEC crossover similar to that of cold Fermi gases, which can be described by the same model, though the interaction must have a finite range, and the crossover is realized by varying the density, as opposed to the scattering length~\cite{Strinati:2018wdg}.} Crucially, the optical trap can be chosen to provide the required harmonic trapping potential for the realization of the state-operator 
correspondence.  Thus, we have an accessible laboratory system that, insofar as finite-charge effects are small, matches our theoretical set-up and invites direct comparison of our predictions to experimental data. One example of a measurable observable is the doubly-integrated axial density, which was determined in~\cite{PhysRevLett.92.120401}.

More in detail, the unitary Fermi gas can be described by a non-relativistic superfluid.
The structure of the effective action was first worked out by Son and Wingate~\cite{Son:2005rv}, up to a set of undetermined Wilsonian coefficients.
The same \ac{eft} was later realized through the technology of the large-charge expansion~\cite{Favrod:2018xov,Kravec:2018qnu,Orlando:2020idm,Hellerman:2021qzz}. To \ac{nlo}\footnote{Sub-leading corrections to the leading-order effective Lagrangian can come from two channels: Derivative corrections and loop corrections.  The former contribute to what are labeled as NLO terms in~\eqref{eq:eff-action-nlo}, while the latter are further suppressed.  Thus, to the order of interest, tree-level analysis will suffice.%
}, in \(3 + 1\) dimensions, the effective Lagrangian with external potential $V(\vect{r})$, takes the form~\cite{Son:2005rv}
\begin{equation}\label{eq:eff-action-nlo}
	\hbar^3 \Lag_{\text{eff}} = c_0m^{3/2}U^{5/2} + c_1 \sqrt{m} U^{-1/2} \del_i U \del^i U + \frac{c_2}{\sqrt{m}} \pqty*{(\del_iD^i\theta)^2 - 9 m \Delta V(\vect{r})}U^{1/2},
\end{equation}
where $\theta$ is the Goldstone of the U(1) superfluid, and
\begin{align}
	U &= D_0 \theta - \tfrac{\hbar}{2m}\del_i\theta \del^i\theta, & D_0 &= \del_t - V(\vect{r}). 
\end{align}
When evaluated on the classical ground state $\theta =\hbar \mu t$, we obtain
\begin{equation}\label{eq:GS-effL}
	\frac{\hbar^3}{m^{3/2}}\Lag_{\text{GS}} = c_0(\hbar\mu - V(\vect{r}))^{5/2} + c_1 \frac{\hbar^2}{m}\frac{(\nabla V(\vect{r}))^2}{\sqrt{\hbar\mu - V(\vect{r})}} - 9 c_2\frac{\hbar^2}{m} \Laplacian V(\vect{r}) \sqrt{\hbar\mu - V(\vect{r})}.
\end{equation}
As an aside, this form of the effective action can be obtained purely from locality and dimensional arguments.  The charge density on the ground state is given by
\begin{equation}
	\frac{\hbar^3}{m^{3/2}}\rho = \frac{5c_0}{2}(\hbar\mu - V(\vect{r}))^{3/2} - \frac{c_1}{2} \frac{\hbar^2}{m}\frac{(\nabla V(\vect{r}))^2}{(\hbar\mu - V(\vect{r}))^{3/2}} - \frac{9c_2}{2} \frac{\hbar^2}{m} \frac{\Laplacian V(\vect{r})}{\sqrt{\hbar\mu - V(\vect{r})}}.
\end{equation}

In the absence of a potential, only the first term in $\Lag_{\text{GS}}$ survives, and the behavior of the Fermi gas is captured by the single Wilsonian coefficient $c_0$. Traditionally, this coefficient is expressed via the so-called Bertsch parameter $\xi$~\cite{Son:2005rv,Chang:2007zzd}, defined as the ratio of the energy density of the unitary Fermi gas to the energy density of a free Fermi gas at the same density, so that
\begin{equation} \label{eq:C0_VS_Bertsch}
	c_0 = \frac{2^{5/2}}{15\pi^2\xi^{3/2}}.
\end{equation}
Once a potential is included, new scales appear and the physics is no longer captured by a single Wilsonian coefficient (equivalently, the Bertsch parameter), as defined above. For the harmonic trap, 
\begin{equation}
	V(\vect{r}) = \tfrac{1}{2}\omega^2 \abs{\vect{r}}^2,
\end{equation}
the bulk contribution to the ground-state energy, respectively the scaling dimension $\Delta$ of the lowest operator with charge $Q$, is given by
\begin{equation}
  \label{eq:trap-energy-Wilson-parameters}
	\frac{\Delta(Q)}{N} = \frac{ E(Q)}{\omega N} = \frac{\sqrt{\xi}}{4}\pqty*{\frac{3Q}{N}}^{4/3} - \sqrt{2}\pi^2 \xi \pqty*{c_1 - \frac{9}{2}c_2}\pqty*{\frac{3Q}{N}}^{2/3} + \dots,
\end{equation}
where $Q$ is the fixed particle number and $N$ is the number of fermion flavors. 

When considering the set-up of a cloud of particles in a harmonic trapping potential, it is obvious that the above \ac{eft} is only valid in the bulk, sufficiently far away from the cloud edge located at a distance \(R_{\text{cl}}\) from the center, where the particle density rapidly drops to zero.
For this reason, the bulk \ac{eft} has to be complemented with terms from the edge~\cite{Hellerman:2020eff,Pellizzani:2021hzx}.
The most general form of terms located at the cloud edge for a system in \(d+1\) dimensions is
\begin{equation}
	Z_{\text{edge}}^{(p)} = \kappa_p \pqty*{\Laplacian^2V(\vect{r}) -\tfrac{1}{d}(\Laplacian\theta)^2}^p \delta(U)(\del_i U)^{(d+4(1-p))/3},
\end{equation}
where $p$ is an integer, $\kappa_p$ is a Wilsonian coefficient, and $\delta(U)$ is an operator-valued delta function that localizes on the cloud edge.

Generally speaking, the contributions of the bulk operators to $\Delta$ can have divergences stemming from the cloud edge if $d$ is even and the operator has positive $Q$-scaling. These divergences can always be regulated, however, by an edge counterterm of the same $\mu$-scaling, leading to $\log(Q)$ terms in $\Delta$~\cite{Hellerman:2020eff,Pellizzani:2021hzx,Hellerman:2021qzz}. 
Additionally, there is a universal $\log(Q)$ term in odd $d$ descending from the Casimir energy of the Goldstone $\theta$. 

The \ac{eft} result for $\Delta$, including contributions from the bulk, the edge, and the Casimir energy, specialized to $3+1$ dimensions, is given by~\cite{Pellizzani:2021hzx,Hellerman:2021qzz}
\begin{equation}
\begin{aligned}
	\Delta(Q)  ={} & Q^{12/9}\left[a_1+\frac{a_2}{Q^{6/9}}+\dots\right]+
	Q^{5/9}\left[b_1+\frac{b_2}{Q^{2/9}}+\frac{b_3}{Q^{4/9}}+\dots\right]\\
	& +\frac{1}{3\sqrt{3}}\log Q + \text{const.} + \dots
\end{aligned}
\end{equation}
Here, the terms starting with $Q^{12/9}$ come from bulk operators, those starting with $Q^{5/9}$ come from edge operators, and $\frac{1}{3\sqrt{3}}\log Q$ is a universal contribution from the Casimir energy.
Further terms are suppressed by negative powers of $Q$.
The coefficients $a_i$, $b_i$ are related to the Wilsonian coefficients and cannot be computed within the framework of the \ac{eft} alone.

\begin{figure}
  \centering
  \includegraphics[width=.7\textwidth]{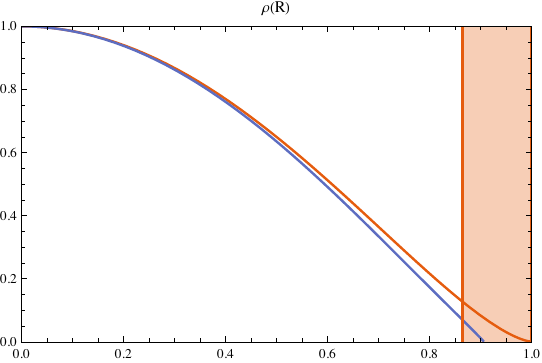}
  \caption{Charge density as a function of the radial position in the harmonic trap for \(\varepsilon = \omega/\mu = 1/20\). Shown are the leading bulk contribution (red) and the bulk contribution including the first correction (blue). The bulk approximation is consistent outside the $\delta$-layer (shaded), where the edge expansion is needed.}
  \label{fig:charge-density}
\end{figure}

\medskip

In this work, we follow a different approach.
We start from a microscopic action, including the trapping potential, and find the form of the effective action in the limit of large $N$ fermion flavors, in a sector of large charge and zero temperature.
As required by locality and the symmetries of the system, we indeed reproduce the form of the ground-state effective action in Eq.~\eqref{eq:GS-effL}.
Importantly, however, the specific microscopic action allows for the direct computation of the Wilsonian coefficients $c_0$, $c_1$ and $c_2$ of the bulk \ac{eft}. 

In principle, can we follow the same strategy as in the relativistic case of the O(N) model at large charge and large N~\cite{Alvarez-Gaume:2019biu}, where standard large-N path integral techniques such as the Stratonovich transform can be employed in a sector of fixed charge. Due to the inhomogeneity introduced by the potential, however, we require more sophisticated techniques. In fact, evaluating the functional determinant from the Gaussian integral of the original fermionic \acp{dof} becomes a nontrivial problem in the presence of the potential. Only when working at large charge is it possible to perform a gradient expansion in the bulk, as the dimensionless parameter $\varepsilon \sim 1/\mu R_{\text{cl}}^2$ in front of the gradient plays the role of $\hbar$ in the quantum mechanical problem of the heat kernel.%

\medskip
At large charge, there is a \ac{ssb}, which, 
at large \(N\), is due to an explicit realization of the Cooper mechanism.  This is because the collective field \(\sigma(x)\) (which acquires a non-trivial \ac{vev} and plays the role of a gap for the fermions) is a bilinear of the fundamental fermions,
\begin{equation}
    \sigma_0(\vect{r}) = \braket{\sigma(\vect{r})} \propto \Braket{\sum_a \psi_{\downarrow a} \psi_{\uparrow a}} \neq 0 .
\end{equation}
Below, we compute the form of the gap to second order in $1/\varepsilon$, finding
\begin{equation}
  \sigma_0(\tau, \vect{r}) = y_0 \pqty{\mu \hbar - V(\vect{r})} + y_1 \frac{\hbar^2}{m} \frac{(\nabla V(\vect{r}))^2}{\pqty{\mu \hbar - V(\vect{r})}^2} + y_2 \frac{\hbar^2}{m} \frac{\Laplacian{V(\vect{r})}}{\mu \hbar - V(\vect{r})} + \dots,
\end{equation}
with $y_0 \approx 1.1622\dots$, $y_1\approx -0.00434691\dots$, and $y_2\approx -0.160794\dots$.

We also compute the energy in the harmonic trap to second order in $1/\varepsilon$, (again, corresponding to the scaling dimension of the lowest operator of charge Q), 
\begin{equation}%
  \frac{\Delta}{N} =  0.8313 \pqty*{\frac{Q}{N} }^{4/3} + 0.26315 \pqty*{\frac{Q}{N} }^{2/3} + \dots
\end{equation}
From there, we can extract the Wilsonian coefficients in the bulk, $c_0, c_1, c_2$,
\begin{align}%
  c_0 &=  0.0841\dots, & c_1 &= 0.006577\dots, & c_2 & \approx 0.004872\dots ,
\end{align}
 and the value of the Bertsch parameter,
 \begin{equation}
    \xi \approx 0.5906\dots .
\end{equation}

This bulk expansion, however, is no longer valid once we approach the cloud edge. Unfortunately, at the edge, we have no small parameter in which to expand the functional determinant (see Figure~\ref{fig:charge-density}). We are, however, able to estimate the scaling of the edge contributions, which indeed matches the \ac{eft} prediction~\cite{Pellizzani:2021hzx,Hellerman:2020eff}.

\medskip
In the previous literature, a number of calculations have been made using various different approximations and approaches. In~\cite{Veillette:2007zz}, a large-N computation was made without an external potential, including subleading corrections in $1/N$. Our leading-order result (the Bertsch parameter) agrees with them. A similar computation has been done in~\cite{Nikolic:2007zz}, which also addresses the case of finite temperature. A mean-field study was performed in~\cite{MANES20091136}. 
A direct comparison of our results can be made with the work of~\cite{Csordas:2010id}, which gives the corrections to gap equation and the charge density in $1/\varepsilon^2$.
We are in complete agreement for the shape of the gap.
There appears to be a discrepancy at \ac{nnlo} between our Wilsonian coefficients and those extracted from their expression for the charge density, however this difference is due to a total derivative term.
There are also numerical lattice~\cite{Forbes:2012yp} studies, though these deal with the opposite regime of small charge, where the quantized nature of the problem is apparent, and it is not clear how to make a direct comparison between our respective results.

The main difference with these previous works is that we use a well-justified approximation, within a controlled perturbation theory, and whose only constraints amount to standard dimensional and symmetry arguments.
This renders a clean, algorithmic way of extracting perturbative data from a very basic set of assumptions. (In fact, the large-charge, large-\(N\) regime justifies the commonly used Thomas--Fermi approximation.)
Moreover, we are able to extract the parametric form of the edge terms from a microscopic description. To our knowledge, this is the first time the droplet edge has been tackled from first principles.

\bigskip

The plan of this article is as follows. In Section~\ref{sec:Model}, we discuss the microscopic model, which is our starting point at large N, and use standard large-N techniques to compute a functional determinant upon performing the Stratonovich transform and integrating out the original fermionic \acp{dof}.
In Section~\ref{sec:Bulk}, we introduce Wigner coordinates and the Moyal product, which allows us to compute the large-charge expansion of the heat kernel in the bulk up to \acs{nnlo}. In Section~\ref{sec:Edge}, we discuss the edge expansion. In Section~\ref{sec:Conclusions}, we give conclusions and an outlook. In the appendices we provide an alternative regularization for the grand potential.

\section{The model}%
\label{sec:Model}

Building upon~\cite{Nikolic:2007zz,Veillette:2007zz,Sachdev2011,Bekaert_2012}, we consider a model of $N$ fermion flavors with two hyperfine ("spin") states $\psi_{\sigma a}$, where $a \in \{1, \ldots, N\}$ and $\sigma \in \{\uparrow, \downarrow\}$, with a unique chemical potential $\mu$, coupled to an external potential $V(\vect{r})$. Working in Euclidean time $\tau$ at zero temperature, \emph{i.e.}, on $\mathbb{R}^4$, this system is described by the action
\begin{equation}
  \label{eq:action}
   	S[\psi] = \int \dd\tau \dd{\vect{r}} \left[
    \bar \psi_{\sigma a} \left( \hbar \partial_\tau{} - \frac{\hbar^2}{2 m} \Laplacian{} - \hbar\mu(\vect{r}) \right) \psi_{\sigma a} + \frac{2 u_0}{N}
    \bar\psi_{\uparrow a} \bar\psi_{\downarrow a} \psi_{\downarrow b} \psi_{\uparrow b} \right],
\end{equation}
where $\bar{\psi}_{\sigma a},~\psi_{\sigma a}$ are Grassmann fields\footnote{$\bar{\psi}_{\sigma a}$ denotes the complex conjugate of $\psi_{\sigma a}$. Note that we work in units where $k_B = 1$, so at finite temperature $T=\beta^{-1}$, these fields would have antiperiodic boundary conditions along the thermal circle $S_{\hbar\beta}^1$, with Matsubara frequencies $\omega_n = (2 n + 1) \pi/(\beta \hbar)$ ($n \in \mathbb{Z}$)~\cite{kapusta_finite-temperature_1989}.} and 
\begin{equation}
	\mu(\vect{r}) \equiv \mu - \frac{1}{\hbar} V(\vect{r}).
\end{equation}
Summation over repeating flavour and spin indices is assumed throughout. We keep the potential generic, as we are interested in computing the Wilsonian coefficients in the \ac{eft}.\footnote{Special forms of $V(\vect r)$ at this stage could lead to accidental symmetries and identifications of operators in the effective action, for instance.}

The bare coupling $u_0<0$ describes an attractive 4-fermion contact interaction which, after renormalization, is proportional to the s-wave scattering length $a_s$. 
The latter is an experimentally tunable parameter, which diverges at the so-called Feshbach resonance. This point in the phase diagram of the Fermi gas is called the \emph{unitary limit}, because the $s$-wave scattering cross-section saturates the bound imposed by unitarity of the S-matrix~\cite{Strinati:2018wdg}. In this limit, the sole physical length scale characterizing the system drops out.  There is strong evidence that this emergent scale invariance is preserved at the quantum level, thereby constituting an example of a strongly interacting nonrelativistic \ac{cft}. 

The action has a manifest $(U(1)\otimes SU(2))^N$ symmetry, where the $U(1)$ copies correspond to particle number conservation of every flavor. However, the complete symmetry group is larger. Indeed, while the kinetic term has an explicit $U(2N)$ symmetry, the interaction term written in the form
\begin{equation}\label{eq:CooperInteraction}
  \frac{u_{0}}{2N} \left( \Psi^{T} \Omega \Psi \right)^{\dagger }  \left( \Psi^{T} \Omega \Psi \right),
\end{equation}
where $\Psi =\left( \psi_{1\uparrow} ...\psi_{N\uparrow } ,\psi_{1\downarrow } ...\psi_{N\downarrow } \right)^{T}$ and \(\Omega = i \sigma_2 \otimes \Id_N\), has a $U(1) \otimes SP(2N, \mathbb{C})$ symmetry. The true symmetry group of the action is thus $U(1) \otimes Sp(2N)$, with the usual definition $Sp(2N) = Sp(2N, \mathbb{C}) \cap U(2N)$~\cite{Bekaert_2012}. Since the Cooper interaction~\eqref{eq:CooperInteraction} appears explicitly in the bare action, we expect \ac{ssb} of the global U(1) symmetry to take place at fixed charge. This is to be contrasted with the Gross--Neveu model~\cite{Dondi:2022zna}, where the breaking is exponentially suppressed (and thus nonperturbative) at large N.

We can introduce a so-called (Hubbard--)Stratonovich field $\sigma(\tau, \vect{r})$ in the Cooper channel (see~\cite{altland_simons_2010} for a discussion of the different channels), namely,
\begin{equation} \label{eq:StrataDef}
    \sigma(\tau, \vect{r}) = -\frac{2 u_0}{N} \sum_{a=1}^N \psi_{\downarrow a}(\tau, \vect{r}) \psi_{\uparrow a}(\tau, \vect{r}),
\end{equation}
resulting in
\begin{equation} \label{eq:Stratonovich-action}
\begin{aligned}
  S[\psi, \sigma]
  & = \int \dd\tau \dd{\vect{r}} \left[
  \bar \psi_{\sigma a} \left( \hbar \partial_\tau - \frac{\hbar^2}{2 m} \Laplacian{} - \hbar \mu(\vect{r}) \right) \psi_{\sigma a} - \frac{N}{2 u_0} \sigma^* \sigma -  %
  \sigma \bar\psi_{\uparrow a} \bar \psi_{\downarrow a} - \sigma^* \psi_{\downarrow a} \psi_{\uparrow a} \right] \\
  & = \int \dd\tau \dd{\vect{r}} \, \left[ -\bar \Psi_a G^{-1}[\sigma] \Psi_a - \frac{N}{2 u_0} \sigma^* \sigma \right],
\end{aligned}
\end{equation}
where, in the second line, we have used the Nambu representation
\begin{equation}
\Psi_a \equiv \begin{pmatrix} \psi_{\uparrow a} \\ \bar\psi_{\downarrow a} \end{pmatrix},
\end{equation}
and $\bar{\Psi }_{a} \equiv \Psi^{\dag}_{a} $, along with the inverse fermion propagator $G^{-1}[\sigma]$,
\begin{equation}
  \label{eq:inverse-propagator}
  G^{-1}[\sigma] \equiv \begin{pmatrix}
                -\hbar \partial_\tau + \frac{\hbar^2}{2 m} \Laplacian{} + \hbar \mu(\tau, \vect{r}) & \sigma(\tau, \vect{r}) \\
                 \sigma(\tau, \vect{r})^* & -\hbar \partial_\tau - \frac{\hbar^2}{2 m} \Laplacian{} - \hbar \mu(\vect{r})
              \end{pmatrix}.
\end{equation}
The Stratonovich field $\sigma$ is a bilinear of the fundamental fermions, which is charged under the $U(1)$. Therefore, if it receives a non-zero \ac{vev}, the symmetry is spontaneously broken. This is the mechanism underlying \ac{bcs} mean-field theory.

With the fermions appearing only quadratically in this formulation, we can integrate them out to find the effective action for \(\sigma\):
\begin{equation}
  \label{eq:sigma-action}
  S[\sigma] = -\hbar N \Tr \log \left( \abs{G^{-1}[\sigma]} \right) - \frac{N}{2u_{0}} \int \dd\tau \dd{\vect{r}} \, \sigma(\tau, \vect{r})^* \sigma (\tau, \vect{r}) .
\end{equation}

As usual in models with a vector symmetry, the fact that the quadratic action for \(\sigma\) is proportional to \(N\) means that we can separate \(\sigma\) into a saddle-point value \(\sigma_0\) plus quantum fluctuations suppressed by \(1/\sqrt{N}\):
\begin{equation}%
  \label{eq:SigmaSaddleAndFluct}
  \sigma(\tau, \vect{r}) = \sigma_{0}(\tau, \vect{r}) +\frac{1}{\sqrt{N}} \hat{\sigma}(\tau,\vect{r}),
\end{equation}
with
\begin{equation}
  \frac{\delta S [\sigma]}{\delta \sigma (\tau, \vect{r})}\Bigg|_{\sigma_0} = 0.
\end{equation}
In \ac{bcs} theory, the saddle $\sigma_0$ is usually referred to as the \emph{gap} (especially in the case without a trap, where it is homogeneous), and the above condition is dubbed the \emph{gap equation}. In practice, we shall solve it to find the $\sigma_0$-profile only at the very end, therefore treating $\sigma_0$ as a generic function.

In Eq.~\eqref{eq:sigma-action}, the leading-order term in a \(1/N\) expansion, namely $S[\sigma] = N \cdot S_0 + \order{N^0}$, is simply given by
 \(S_0 \equiv \frac{1}{N} S[\sigma_0]\). Similarly, we shall use \(G_0 \equiv G[\sigma_0]\). With the external potential \(V(\vect{r})\) time-independent, we can write the inverse fermion propagator as the sum of a time-derivative and a position-dependent operator $B(\vect{r})$, which we shall refer to as the \ac{bdg} operator:
\begin{align}\label{eq:BdG}
    G_0^{-1} = -\hbar \Id \cdot \del_\tau{} + B(\vect{r}),
   &&\text{where} && 
    B(\vect{r}) \equiv
    \begin{pmatrix}
        -h(\vect{r}) & \sigma_0(\vect{r}) \\
        \sigma_0(\vect{r})^* & h(\vect{r})
    \end{pmatrix},
\end{align}
and \(h(\vect{r})\) is the one-particle Hamiltonian 
\begin{equation}
	h(\vect{r}) = -\frac{\hbar^2}{2m} \Laplacian{} - \hbar \mu(\vect{r}).
\end{equation}
Since $B(\vect{r})$ is Hermitian, its eigenvalues are real and come in pairs. Moreover, the eigenvalues of the piece proportional to the identity in $G_0^{-1}$ are given by $i \hbar \omega_n$, where $\omega_n$ ($n \in \mathbb{Z}$) are the Matsubara frequencies (see previous footnote). It is possible to perform the sum over them explicitly and, in the zero-temperature limit, the result takes the simple form
\begin{equation} \label{eq:ZeroT_TrLog}
    \lim_{\beta\to\infty} \Tr \log\left( \abs{G_0^{-1}} \right) = \frac{\beta}{2} \Tr(\abs{B}) .
\end{equation}
The factor \(1/2\) in front of the trace has a direct physical interpretation, since we must distinguish between the ``large'' first-quantized Hilbert space, generated by all the modes (which is \emph{not} a quantum system) and the ``physical'' Hilbert space of positive-energy modes.%

One can obtain Eq.~\eqref{eq:ZeroT_TrLog} by introducing the spectral zeta function associated with an operator $\mathcal{O}$:
\begin{equation}
    \zeta_\mathcal{O}(s) \equiv \Tr*( \abs{\mathcal{O}}^{-s}) = \sum_{\alpha} \abs{o_\alpha}^{-s},
\end{equation}
where the eigenvalues $o_\alpha$ of $\mathcal{O}$ are labelled by a generic index $\alpha$. With this, it is easy to show that
\begin{equation}
    \lim_{\beta\to\infty} \zeta_{G_0^{-1}}(s)
    = \frac{\beta \Gammaop*( \frac{s-1}{2} )}{2 \sqrt{\pi} \Gammaop*( \frac{s}{2} )} \zeta_B(s-1)
    = \frac{\beta \Gammaop*( \frac{s-1}{2} )}{2 \sqrt{\pi} \Gammaop*( \frac{s}{2} )} \zeta_{B^2}\left( \frac{s-1}{2} \right),
\end{equation}
so that
\begin{equation}
    \lim_{\beta\to\infty} \Tr \log*( \abs{G_0^{-1}} ) = -\lim_{\beta\to\infty} \eval*{ \frac{d}{d s} \zeta_{G_0^{-1}}(s) }_{0^+}  = \frac{\beta}{2} \zeta_B(-1) = \frac{\beta}{2} \zeta_{B^2}\left( -\frac{1}{2} \right).
\end{equation}
For computational convenience, the last expressions above are cast in terms of the positive-definite operator \(B^2\). In order to compute such an object, it is useful to set up the heat kernel problem associated with $B^2$. The heat kernel $K_{B^2}(\vect{r}_1, \vect{r}_2; \proptime)$ is a 2-by-2 matrix satisfying
\footnote{As usual, one should not confuse the proper time \(\proptime\) with the physical time.}
\begin{equation} \label{eq:HeatK_Problem}
\begin{dcases}
 \left( \del_{\proptime}{} + B^2(\vect{r}_1) \right) K_{B^2}(\vect{r}_1, \vect{r}_2; \proptime), = 0\\
     \lim_{\proptime\to0^+} K_{B^2}(\vect{r}_1, \vect{r}_2; \proptime) =\deltaop(\vect{r}_1 - \vect{r}_2) \cdot \Id_2 ,
\end{dcases}
\end{equation}
which formally means that $K_{B^2}(\vect{r}_1, \vect{r}_2; \proptime) = \braket{\vect{r}_1| e^{-B^2 \proptime} |\vect{r}_2}$. We denote the Dirac $\delta$-function with a hat to avoid later confusion, and $\Id_2$ is the 2-by-2 identity matrix. In turn, the coincident-point limit of the heat kernel allows one to compute the zeta function associated with $B^2$ through its Mellin transform
\begin{equation} \label{eq:ZetaMellin}
  \zeta_{B^2}(s)
  = \frac{1}{\Gamma(s)} \int_0^\infty \frac{\dd{\proptime}}{\proptime} \proptime^s \int \dd{\vect{r}} \Tr\left(K_{B^2}(\vect{r}, \vect{r}; \proptime)\right),
\end{equation}
where $\Tr$ now just stands for the matrix trace.

Effectively, this turns the computation of the one-loop determinant --- namely, the $\Tr\log(\abs{G_0^{-1}})$ term --- into the quantum mechanical problem of a particle with Hamiltonian \(B^2\). We will see in the following section that this is a good starting point for evaluating the functional determinant perturbatively in the limit of large chemical potential (or, equivalently, large charge), even when the system is inhomogeneous due to the presence of an external potential.

Before proceeding, it is useful to reformulate Eqs.~\eqref{eq:HeatK_Problem} in dimensionless quantities, which will generically be denoted with a bar. As mentioned, we will eventually solve this problem perturbatively, but we need not commit to any approximation at this stage. The goal at this point is rather to express the problem in an exact form suggestive of a perturbative resolution. 
Consider the case of a generic potential $V(\vect{r})$ that vanishes at the origin and that (classically) confines the particles to a finite region of space. Correspondingly, let $R_{cl}$ be the length scale associated with the smallest distance from the origin such that $\mu(\vect{r}) = 0$, \emph{i.e.}, where the particle density vanishes classically. For concreteness (and since in the end we want to compute conformal dimensions), one can consider the harmonic potential\footnote{Though we do not need to commit to this choice for now.}
$V(\vect{r}) = \frac{m \omega^2}{2} \vect{r}^2$, in which case,
\begin{equation} \label{eq:R_cl}
    R_{cl} \equiv \sqrt{\frac{2 \hbar \mu}{m \omega^2}}.
\end{equation}
This length scale allows us to work in terms of dimensionless coordinates defined as
\begin{equation}
    \bar{\vect{r}} \equiv \frac{\vect{r}}{R_{cl}}.
\end{equation}
Moreover, upon defining $\bar B(\bar{\vect{r}}) \equiv \frac{1}{\hbar \mu} B(\vect{r})$ and $K_{\bar B^2}(\bar{\vect{r}}_1, \bar{\vect{r}}_2; \bar\proptime) \equiv R_{cl}^3 K_{B^2}(\vect{r}_1, \vect{r}_2; \proptime)$ with $\bar\proptime \equiv (\hbar \mu)^2 \proptime$, we obtain
\begin{equation} \label{eq:HeatK_Problem_DimLess}
\begin{dcases}
     \left( \del_{\bar\proptime}{} + \bar B^2(\bar{\vect{r}}_1) \right) K_{\bar B^2}(\bar{\vect{r}}_1, \bar{\vect{r}}_2; \bar\proptime) = 0, \\
    \lim_{\bar\proptime\to0^+} K_{\bar B^2}(\bar{\vect{r}}_1, \bar{\vect{r}}_2; \bar\proptime) = \deltaop(\bar{\vect{r}}_1 - \bar{\vect{r}}_2) \cdot \Id_2 ,
\end{dcases}
\end{equation}
where everything is dimensionless. Note that, explicitly, we have
\begin{equation}
    \bar B(\bar{\vect{r}}) =
    \begin{pmatrix}
        -\bar h(\bar{\vect{r}}) & \bar\sigma(\bar{\vect{r}}) \\
        \bar\sigma(\bar{\vect{r}})^* & \bar h(\bar{\vect{r}})
    \end{pmatrix},
\end{equation}
with $\bar\sigma(\bar{\vect{r}}) \equiv \frac{\sigma_0(\vect{r})}{\hbar \mu}$, $\bar h(\bar{\vect{r}}) \equiv \frac{h(\vect{r})}{\hbar \mu} = -\frac{\hbar}{2 m \mu R_{cl}^2} \Laplacian_{\bar{\vect{r}}} + \bar V(\bar{\vect{r}}) - 1$ and $\bar V(\bar{\vect{r}}) \equiv \frac{1}{\hbar \mu} V(\vect{r})$. In the particular case of the harmonic potential, we simply have $\bar V(\bar{\vect{r}}) = \bar{\vect{r}}^2$.

\section{The bulk expansion}%
\label{sec:Bulk}

We have seen that the computation of the effective action at leading order in $N$ reduces to evaluating the trace of the absolute value (or the square) of the \ac{bdg} operator (cf.~Eq.~\eqref{eq:ZeroT_TrLog}).
In the absence of a confining potential, this is a standard calculation~\cite{Nikolic:2007zz,Veillette:2007zz}. 
In the presence of a confining potential, the problem is no longer translationally invariant and the particles are confined to a spherical region at the edge of which the particle density drops to zero.

In this section, we show how to perturbatively compute the free energy in the presence of a potential when the particle number is large, thereby providing the first explicit verification of the predictions of nonrelativistic large-charge \ac{eft}.
The underlying controlling parameter of this perturbative computation is the particle density, which naturally breaks down close to the edge of the particle cloud.
However, \ac{eft} considerations not only tell us that this had to be anticipated, but also that ``exotic'' contributions enter the large-charge expansion of the free energy. While the setup in this section is not suited to capture these contributions, we discuss in Section~\ref{sec:Edge} how to perform the matching between the expansion in the bulk of the cloud with the solution close to the edge, and find further agreement with the \ac{eft} predictions.

\subsection{Wigner coordinates and Moyal product}
\label{sec:wigner-moyal}

The lack of translational invariance can be addressed through the introduction of so-called mixed, or Wigner, coordinates~\cite{Wigner:1932eb,rammer_2007}.
The idea is to first write \(\bar B(\bar{\vect{r}})\) as a bilocal operator via
\begin{equation}
    \bar B(\bar{\vect{r}}_1, \bar{\vect{r}}_2) \equiv \deltaop(\bar{\vect{r}}_1 - \bar{\vect{r}}_2) \bar B(\bar{\vect{r}}_1) = \begin{pmatrix}
        -\bar h(\bar{\vect{r}}_1, \bar{\vect{r}}_2) & \bar\sigma(\bar{\vect{r}}_1) \deltaop(\bar{\vect{r}}_1 - \bar{\vect{r}}_2) \\
        \bar\sigma(\bar{\vect{r}}_1)^* \deltaop(\bar{\vect{r}}_1 - \bar{\vect{r}}_2) & \bar h(\bar{\vect{r}}_1, \bar{\vect{r}}_2)
    \end{pmatrix},
\end{equation}
where
\begin{equation}
    \bar h(\bar{\vect{r}}_1, \bar{\vect{r}}_2) \equiv -\frac{\hbar}{2 m \mu R_{cl}^2} \left( \Laplacian_{\bar{\vect{r}}_1}  \deltaop(\bar{\vect{r}}_1 - \bar{\vect{r}}_2) \right) + \left( \bar V(\bar{\vect{r}}_1) - 1 \right) \deltaop(\bar{\vect{r}}_1 - \bar{\vect{r}}_2), 
\end{equation}
and then perform a Fourier transform.  Introducing relative and center-of-mass positions
\begin{equation}
\begin{cases}
    \bar{\vect{r}}_{ij} \equiv \bar{\vect{r}}_i - \bar{\vect{r}}_j, \\
    \bar{\vect{R}}_{ij} \equiv \frac{\bar{\vect{r}}_i + \bar{\vect{r}}_j}{2},
\end{cases}
\end{equation}
a bilocal Fourier transform of a function $A(\bar{\vect{r}}_1, \bar{\vect{r}}_2)$ can be defined as
\begin{equation}
    a(\bar{\vect{R}}, \bar{\vect{p}}) \equiv \int \dd{\bar{\vect{r}}} e^{-\frac{i\bar{\vect{p}}\cdot\bar{\vect{r}}}{\varepsilon}} A\left(\bar{\vect{R}} + \frac{\bar{\vect{r}}}{2}, \bar{\vect{R}} - \frac{\bar{\vect{r}}}{2}\right),
\end{equation}
and the inverse transformation is
\begin{equation}
    A(\bar{\vect{r}}_i, \bar{\vect{r}}_j) = \int \frac{\dd{\bar{\vect{p}}}}{(2 \pi \varepsilon)^3} e^{\frac{i\bar{\vect{p}}\cdot\bar{\vect{r}}_{ij}}{\varepsilon}} a(\bar{\vect{R}}_{ij}, \bar{\vect{p}}).
\end{equation}
We choose to put a bar on the momentum as well to indicate that it is dimensionless. In some sense, this transform allows one to disentangle the ``microscopic'' dynamics of the system (associated with $\bar{\vect{r}}_{ij}$) from the ``macroscopic'' properties (associated with $\bar{\vect{R}}_{ij}$) resulting from the external potential~\cite{rammer_2007}. 
For the moment, $\varepsilon$ is an arbitrary real parameter which will be assigned physical meaning soon. 

From this definition of the Fourier transform, it is easy to show the following general statement~\cite{Groenewold:1946kp,Moyal:1949sk}. If $C(\bar{\vect{r}}_1, \bar{\vect{r}}_2)$ is related to two other bilocal functions $A$ and $B$ by
\begin{equation}
	C(\bar{\vect{r}}_1, \bar{\vect{r}}_3) = \int \dd{\bar{\vect{r}}_2} A(\bar{\vect{r}}_1, \bar{\vect{r}}_2) B(\bar{\vect{r}}_2, \bar{\vect{r}}_3),
\end{equation}
then
\begin{equation}
	c(\bar{\vect{R}}, \bar{\vect{p}}) = a(\bar{\vect{R}}, \bar{\vect{p}}) \star b(\bar{\vect{R}}, \bar{\vect{p}}),
\end{equation}
where $\star$ is the Moyal product, defined by
\begin{equation}
\begin{aligned}
    a(\bar{\vect{R}}, \bar{\vect{p}}) \star b(\bar{\vect{R}}, \bar{\vect{p}})
    & \equiv a(\bar{\vect{R}}, \bar{\vect{p}}) \exp*[\frac{i \varepsilon}{2} \pqty*{\cev{\del}_{\bar{\vect{R}}} \vec{\del}_{\bar{\vect{p}}} - \cev{\del}_{\bar{\vect{p}}} \vec{\del}_{\bar{\vect{R}}}}] b(\bar{\vect{R}}, \bar{\vect{p}}) \\
    & = \sum_{k=0}^\infty \left( \frac{i \varepsilon}{2} \right)^k \frac{1}{k!} \poisson*{a(\bar{\vect{R}}, \bar{\vect{p}}), b(\bar{\vect{R}}, \bar{\vect{p}})}_k,
\end{aligned}
\end{equation}
with what we shall call the \emph{$k$-Poisson bracket}:
\begin{equation} \label{eq:KPoisson_def}
    \poisson*{a(\bar{\vect{R}}, \bar{\vect{p}}), b(\bar{\vect{R}}, \bar{\vect{p}})}_k \equiv a(\bar{\vect{R}}, \bar{\vect{p}}) \pqty*{\cev{\del}_{\bar{\vect{R}}} \vec{\del}_{\bar{\vect{p}}} - \cev{\del}_{\bar{\vect{p}}} \vec{\del}_{\bar{\vect{R}}}}^k b(\bar{\vect{R}}, \bar{\vect{p}}).
\end{equation}
Of course, the $k=1$ case is just the normal Poisson bracket.\footnote{
Explicit computations involved in this paper contain at most the $k=2$ case, which can also be put in the following form. Given the Poisson bivector $\Pi = i\sigma_2 \otimes \Id_d$ ($d = 3$ in our case) and the derivative $\del_I$ ($I = 1, \ldots, 2d$) acting on phase-space coordinates $(\bar{\vect{R}}, \bar{\vect{p}})$, and $\del_{IJ} \equiv \del_I \del_J$, we have
\begin{align}
  \poisson*{a, b}_1 &=  \Pi^{IJ} \del_I a \, \del_J b , \\
  \poisson*{a, b}_2 &=  \Pi^{IJ}\Pi^{LM}\del_{IL} a \, \del_{JM} b.
\end{align}
}

This formalism was originally introduced to describe quantum mechanics in phase space, and is adapted to solve our heat kernel problem, which takes the form
\begin{equation}
\begin{dcases}
	 \del_{\bar\proptime} k(\bar{\vect{R}}, \bar{\vect{p}}; \bar\proptime) + b(\bar{\vect{R}}, \bar{\vect{p}}; \proptime) \star b(\bar{\vect{R}}, \bar{\vect{p}}; \bar\proptime) \star k(\bar{\vect{R}}, \bar{\vect{p}}; \proptime) = 0, \\
    \lim_{\bar\proptime\to0^+} k(\bar{\vect{R}}, \bar{\vect{p}}; \bar\proptime) = \Id_2  ,
\end{dcases}
\end{equation}
where $k(\bar{\vect{R}}, \bar{\vect{p}}; \bar\proptime)$ denotes the Fourier transform of $K_{\bar B^2}(\bar{\vect{r}}_1, \bar{\vect{r}}_2; \bar\proptime)$, as defined above, and 
\begin{equation}
    b(\bar{\vect{R}}, \bar{\vect{p}}) =
    \begin{pmatrix}
        -\bar h(\bar{\vect{R}}, \bar{\vect{p}}) & \bar\sigma(\bar{\vect{R}}) \\
        \bar\sigma(\bar{\vect{R}})^* & \bar h(\bar{\vect{R}}, \bar{\vect{p}})
    \end{pmatrix},
\end{equation}
with the dimensionless phase-space Hamiltonian given by
\begin{equation} \label{eq:Hamiltonian_PhaseSpace}
    \bar h(\bar{\vect{R}}, \bar{\vect{p}}) = \frac{\hbar}{2 m \mu R_{cl}^2 \varepsilon^2} \bar{\vect{p}}^2 + \bar V(\bar{\vect{R}}) - 1.
\end{equation}
Note that the gap \(\bar\sigma\) is only a function of \(\bar{\vect{R}}\) because it encodes pointwise interactions.

Since, at leading order in $N$, we can identify $\frac{1}{\hbar\beta} S_0$ with the grand-canonical potential $\Omega$, its zero-temperature limit can be expressed in terms of the solution $k(\bar{\vect{R}}, \bar{\vect{p}}; \bar\proptime)$ to the above heat kernel problem as %
\begin{equation} \label{eq:S0_DimLessStuff}
\begin{aligned}
    \Omega(\mu)
    = -\frac{\hbar \mu}{2} \left[ \int_0^\infty \frac{\dd{\bar\proptime}}{\Gamma(s) \bar\proptime} \bar\proptime^s \int \frac{\dd{\bar{\vect{R}}} \dd{\bar{\vect{p}}}}{(2 \pi \varepsilon)^3} \Tr\left(k(\bar{\vect{R}}, \bar{\vect{p}}; \bar\proptime)\right) \right]_{s=-\frac{1}{2}} - \frac{(\hbar \mu)^2 R_{cl}^3}{2 u_0} \int \dd{\bar{\vect{R}}} \abs{\bar\sigma(\bar{\vect{R}})}^2,
\end{aligned}
\end{equation}
on top of which one needs to impose the gap equation. The charge is then given by
\begin{equation}
    Q = -\frac{1}{\hbar} \frac{\del\Omega(\mu)}{\del\mu},
\end{equation}
which can be inverted to find $\mu$ as a function of $Q$, and the (zero-temperature) free energy is
\begin{equation}
    F(Q) = \Omega(\mu(Q)) + \hbar \mu(Q) \cdot Q.
\end{equation}
In the specific case of a critical system in a harmonic trap $V(\vect{r}) = \frac{m \omega^2}{2} \vect{r}^2$ (\emph{i.e.}, $\bar V(\bar{\vect{R}}) = \bar{\vect{R}}^2$ and $\varepsilon = {\omega}/({2 \mu})$, cf.~Eq.~\eqref{eq:R_cl}), the free energy is related to the conformal dimension of the lightest operator of charge $Q$ as~\cite{Nishida:2007pj,Goldberger:2014hca}
\begin{equation} \label{eq:DeltaQ_vs_FreeEnergy}
    \Delta(Q) = \frac{1}{\hbar\omega} F^\text{crit.}(Q).
\end{equation}

\subsection{Large-charge expansion of the heat kernel}%
\label{sec:large-charge-BdG}

Within the large-$N$ expansion, this general construction is thus far exact, although the fact that the Moyal product can be expanded as an asymptotic series in $\varepsilon$ suggests a perturbative calculation of the heat kernel problem.
It is convenient to canonically normalize the momentum in the phase-space Hamiltonian in Eq.~(\ref{eq:Hamiltonian_PhaseSpace}) by fixing
\begin{equation} \label{eq:EpsilonFixed}
    \varepsilon = \sqrt{\frac{\hbar}{2 m \mu}} \frac{1}{R_{cl}},
\end{equation}
which is small if the (dominating) scale associated with the potential is smaller than the scale set by the charge density. This is the large-charge limit because, as we will show, $Q \sim \varepsilon^{-3}$. Our goal is to compute $\Omega(\mu)$ as an expansion in $\varepsilon$,
\begin{equation} \label{eq:S0_EpsilonExpand}
    \Omega(\mu) \equiv \Omega_{LO}(\mu) + \varepsilon \cdot \Omega_{NLO}(\mu) + \varepsilon^2 \cdot \Omega_{NNLO}(\mu) + \cdots,
\end{equation}
up to quadratic order, and thus to compute the corresponding free energy to this order.\footnote{As a reminder, the expansion to this order covers the dynamics of the effective action~\eqref{eq:eff-action-nlo} to 'NLO', in the language of~\cite{Son:2005rv}. On general grounds we expect the linear piece in \(\varepsilon\) to vanish because \(\Omega(\mu)\) is the the saddle value of the function \(\Omega(\sigma)\), and \(\Omega_{NLO}(\mu) \) is the value of its first variation.}

This amounts to a semiclassical analysis of the Hamiltonian, Eq.~\eqref{eq:Hamiltonian_PhaseSpace}, which is consistent as long as the potential term is bigger than the kinetic term.
In the bulk region, the gap profile \(\sigma(\vect{r})\) does not vary rapidly on the scale defined by its own \ac{vev}:
\begin{align}
  \label{eq:bulk-condition}
  \sigma(\vect{r}) \gg \frac{\hbar^2}{2 m}\frac{\pqty*{\del_{\vect{r}} \sigma(\vect{r})}^2}{\sigma(\vect{r})^2} . && \text{(bulk condition).}
\end{align}
In terms of dimensionless variables, this turns into
\begin{equation}
  \bar \sigma(\bar{\vect{R}}) \gg \epsilon^2 \frac{\pqty*{\del_{\bar{\vect{R}}} \bar \sigma(\bar{\vect{R}})}^2}{\bar \sigma(\bar{\vect{R}})^2} ,
\end{equation}
which is satisfied as long as \(\bar \sigma(\bar{\vect{R}})\) and its derivative with respect to \(\bar{\vect{R}}\) are both of order one.
More precisely, this means that $1 - \bar V (\bar{\vect{R}})$ must be bigger than some new parameter $\delta$, which we will evaluate below.
The construction is analogous to the standard \ac{wkb} approximation, in which the expansion in $\hbar$ is valid in the bulk away from the turning points.
For the moment, we concentrate on the bulk region, where $1 - \bar V (\bar{\vect{R}}) > \delta$ (in boundary layer theory this is called the ``outer region'' \cite{bender1999advanced}).
We will turn to the edge of the cloud (the ``inner region''), which necessitates a separate treatment, in the next section.

Having clarified the interval of validity of our approximation, we can use standard phase-space quantum mechanics. With our choice of $\varepsilon$~\eqref{eq:EpsilonFixed}, the one-particle Hamiltonian in phase space takes the form 
\begin{equation} \label{eq:Hamiltonian_PhaseSpaceBis}
    \bar h(\bar{\vect{R}}, \bar{\vect{p}}) = \bar{\vect{p}}^2 + \bar V(\bar{\vect{R}}) - 1.
\end{equation}
The dependence on $\varepsilon$ has been reabsorbed by the Fourier transform in the same manner that $\hbar$ does not appear in the phase-space Hamiltonian in quantum mechanics.
This expression shows explicitly how in Wigner coordinates the contribution of the external potential, which depends only on the center-of-mass coordinate \(\bar{\vect{R}}\), is separated from the kinetic part depending only on the (dual) relative coordinate \(\bar{\vect{p}}\).

In the large-charge regime $\varepsilon \ll 1$, it is then natural to make the following Ansätze for the gap and the heat kernel:
\begin{align} \label{eq:gap-expansion}
	\bar \sigma(\bar{\vect{R}}) & \equiv \sum_{k=0}^\infty \varepsilon^k \Sigma_k(\bar{\vect{R}}), \\
	  k(\bar{\vect{R}}, \bar{\vect{p}}; \bar\proptime) & 
    \equiv \sum_{j=0}^\infty \varepsilon^j
    \begin{pmatrix}
        \Kdiag_j(\bar{\vect{R}}, \bar{\vect{p}}; \bar\proptime) & \Koff_j(\bar{\vect{R}}, \bar{\vect{p}}; \bar\proptime) \\
        \Koff_j(\bar{\vect{R}}, \bar{\vect{p}}; \bar\proptime)^* & \Kdiag_j(\bar{\vect{R}}, \bar{\vect{p}}; \bar\proptime)
    \end{pmatrix}.
\end{align}
Note that the matrix trace $\Tr \left( k(\bar{\vect{R}}, \bar{\vect{p}}; \bar\proptime) \right) = 2 \cdot \sum_{j=0}^\infty \varepsilon^j \Kdiag_j(\bar{\vect{R}}, \bar{\vect{p}}; \bar\proptime)$ will bring a factor of 2 into Eq.~\eqref{eq:S0_DimLessStuff}.

In the following, we will need to solve the saddle-point condition \(\delta \Omega / \delta \sigma = 0\).
It is convenient to reformulate it in terms of \(\Sigma_0\) by the chain rule
\begin{equation}
  \frac{\delta}{\delta \Sigma_0} = \frac{\delta \sigma}{\delta \Sigma_0} \frac{\delta}{\delta \sigma} = \frac{\delta \bar \sigma}{\delta \Sigma_0} \frac{\delta \sigma}{\delta \bar \sigma} \frac{\delta}{\delta \sigma} = \hbar \mu \frac{\delta}{\delta \sigma} ,
\end{equation}
so that the saddle-point condition reads \(\delta \Omega / \delta \Sigma_0 = 0\).
All the information about the saddle (\emph{i.e.}, the values of the functions \(\Sigma_j\)) is contained in the variations with respect to \(\Sigma_0\).
The same information is also contained in the variations with respect to any of the \(\Sigma_j\), but this is suppressed in the \(\varepsilon\) expansion because
\begin{equation}
  \frac{\delta}{\delta \Sigma_j} = \frac{\delta \sigma}{\delta \Sigma_j} \frac{\delta}{\delta \sigma} = \frac{\delta \bar \sigma}{\delta \Sigma_j} \frac{\delta \sigma}{\delta \bar \sigma} \frac{\delta}{\delta \sigma} = \hbar \mu \varepsilon^j \frac{\delta}{\delta \sigma} .
\end{equation}
In the expansion of \(\Omega\), the function  \(\Sigma_j\) appears linearly at order \(\varepsilon^j\):
\begin{multline}
  \Omega(\bar \sigma) = \Omega_{LO}(\Sigma_0) + \pqty*{ \Omega_{NLO}(\Sigma_0) +  \Sigma_1 \hat \Omega_{NLO}(\Sigma_0) } \varepsilon \\
  +  \pqty*{ \Omega_{NNLO}(\Sigma_0, \Sigma_1) +  \Sigma_2 \hat \Omega_{NNLO}(\Sigma_0)} \varepsilon^2 + \dots,
\end{multline}
and, by the equation above, \( \hat \Omega_{NLO}(\Sigma_0) = \Omega_{LO}'(\Sigma_0) \), which vanishes at the saddle.
It follows that, apart from the leading order, the saddle-point value  \(\braket{\Omega}\) at order \(j\) does not depend on the saddle value \(\braket{\Sigma_j}\).

The phase-space Hamiltonian is quadratic in the variables $\bar{\vect{R}}$ and $\bar{\vect{p}}$, and it is easy to express the product \(b(\bar{\vect{R}}, \bar{\vect{p}}) \star b(\bar{\vect{R}}, \bar{\vect{p}})\) in closed form,
\begin{equation}
\begin{aligned}
    b(\bar{\vect{R}}, \bar{\vect{p}}) \star b(\bar{\vect{R}}, \bar{\vect{p}})
    & =
    \begin{pmatrix}
        h(\bar{\vect{R}}, \bar{\vect{p}})^2 + \abs{\bar\sigma(\bar{\vect{R}})}^2 - \frac{\varepsilon^2}{2} \Laplacian_{\bar{\vect{R}}}(\bar V(\bar{\vect{R}}))
        & \kern-10pt 2 i \varepsilon \bar{\vect{p}} \cdot \nabla_{\bar{\vect{R}}} \bar \sigma(\bar{\vect{R}})
        \\
        -2i \varepsilon \bar{\vect{p}} \cdot \nabla_{\bar{\vect{R}}} \bar \sigma(\bar{\vect{R}})
        & \kern-10pt h(\bar{\vect{R}}, \bar{\vect{p}})^2 + \abs{\bar\sigma(\bar{\vect{R}})}^2 - \frac{\varepsilon^2}{2} \Laplacian_{\bar{\vect{R}}}(\bar V(\bar{\vect{R}}))
  \end{pmatrix}  \\
  & \equiv
  \sum_{i=0}^\infty \varepsilon^i
  \begin{pmatrix} 
        \bbdiag_i(\bar{\vect{R}}, \bar{\vect{p}}) & \bboff_i(\bar{\vect{R}}, \bar{\vect{p}}) \\
        \bboff_i^*(\bar{\vect{R}}, \bar{\vect{p}}) & \bbdiag_i(\bar{\vect{R}}, \bar{\vect{p}})
  \end{pmatrix},
\end{aligned} 
\end{equation}
where we have identified, order by order,
\begin{equation}
\begin{dcases}
    \bbdiag_0(\bar{\vect{R}}, \bar{\vect{p}})  \equiv h(\bar{\vect{R}}, \bar{\vect{p}})^2 + \abs{\Sigma_0(\bar{\vect{R}})}^2, \\
    \bbdiag_{i}(\bar{\vect{R}}, \bar{\vect{p}})  \equiv \sum_{k=0}^i \Sigma_k^*(\bar{\vect{R}}) \Sigma_{i-k}(\bar{\vect{R}}) - \frac{\delta_{i2}}{2} \Laplacian_{\bar{\vect{R}}}(\bar V(\bar{\vect{R}})) & \text{for $i = 1,2,\dots $,}
\end{dcases}
\end{equation}
and
\begin{equation}
\begin{cases}
    \bboff_0(\bar{\vect{R}}, \bar{\vect{p}}) \equiv 0, \\
    \bboff_i(\bar{\vect{R}}, \bar{\vect{p}})  \equiv 2i \bar{\vect{p}} \cdot \nabla_{\bar{\vect{R}}} \Sigma_{i-1}(\bar{\vect{R}}) & \text{for $i=1,2,\dots$.}
\end{cases}
\end{equation}
With this, the heat kernel equation can be understood hierarchically. Expanding the Moyal product in powers of $\varepsilon$ with the $k$-Poisson bracket notation introduced in Eq.~\eqref{eq:KPoisson_def}, and dropping the arguments to avoid cluttering the notation, the order-$\varepsilon^n$ heat kernel problem becomes
\begin{equation} \label{eq:HeatK_Problem_OrderN}
\begin{dcases}
 \del_{\bar\proptime} \Kdiag_n + \sum_{j=0}^n \sum_{k=0}^{n-j} \frac{i^k}{2^k k!} \left[ \poisson*{\bbdiag_{n-j-k}, \Kdiag_j}_k + \poisson*{\bboff_{n-j-k}, \Koff_j^*}_k \right] = 0, \\
\del_{\bar\proptime} \Koff_n + \sum_{j=0}^n \sum_{k=0}^{n-j} \frac{i^k}{2^k k!} \left[ \poisson*{\bbdiag_{n-j-k}, \Koff_j}_k + \poisson*{\bboff_{n-j-k}, \Kdiag_j}_k \right] = 0,
\end{dcases}
\end{equation}
with \ac{ic} $\lim\limits_{\bar\proptime\to 0} \Kdiag_n = \delta_{0n}$ and $\lim\limits_{\bar\proptime\to 0} \Koff_n = 0$. Note that each line contains a finite number of contributions and, for a given $n$, the two equations are decoupled, since $\bboff_0 = 0$. We investigate the order $\varepsilon^0$ contributions in the next section, and we will elaborate further on this system of equations when we reach subleading contributions.

\subsection{Leading order}%
\label{sec:leading-order-bulk}

At \ac{lo}, the Moyal product is just a pointwise product.
This means that, at this order, $\bar\sigma$ commutes with the momentum and may therefore be treated as effectively constant.
In other words, for very large $\mu$, the position-dependent $\mu(\vect{r})$ can be regarded as a slowly varying function, and the computation is formally the same as the one without potential.
In particular, by dimensional analysis and locality, the gap \(\Sigma_0(\bar{\vect{R}})\) must be proportional to the effective chemical potential $\mu(\vect{r})$, namely
\begin{equation} \label{eq:Sigma0_LO}
   \Sigma_0(\bar{\vect{R}}) = y_0 \pqty*{ 1 - \bar V(\bar{\vect{R}})},
   \quad \text{with} \quad y_0 \in \mathbb{R},
\end{equation}
and we must thereby reproduce the standard mean-field result for the Bertsch parameter in the absence of an external potential~\cite{Eagles:1969zz} (see also, \emph{e.g.},~\cite{Veillette:2007zz}).
It should soon become clear that this arises naturally in the \ac{lo} analysis.

The \ac{lo} contributions are contained in the $n=0$ (and therefore $j = k = 0$) terms in Eq.~\eqref{eq:HeatK_Problem_OrderN}. The heat kernel equations at this order are
\begin{equation}
  \begin{cases}
    \del_{\bar\proptime} \Kdiag_0 + \bbdiag_0 \Kdiag_0 = 0, \\
    \del_{\bar\proptime} \Koff_0 + \bbdiag_0 \Koff_0 = 0 ,
  \end{cases}
\end{equation}
with \ac{ic} $\Kdiag_0(\bar{\vect{R}},\bar{\vect{p}}; 0) = 1$ and $\Koff_0(\bar{\vect{R}},\bar{\vect{p}}; 0) = 0$. Thus, 
$\Kdiag_0 = e^{-\bbdiag_0 \bar\proptime}$ and $\Koff_0 = 0$,
and Eq.~\eqref{eq:S0_DimLessStuff} at \ac{lo} becomes
\begin{equation}
  \begin{aligned}
    \Omega_{LO}(\mu)
    ={}& -\hbar \mu \left[ \int_0^\infty \frac{\dd{\bar\proptime}}{\Gamma(s) \bar\proptime} \bar\proptime^s \int \frac{\dd{\bar{\vect{R}}} \dd{\bar{\vect{p}}}}{(2\pi \varepsilon)^3} e^{ -\pqty*{\pqty*{ 1 - \bar V(\bar{\vect{R}}) -  \bar{\vect{p}}^2 }^2 + \abs{\Sigma_0(\bar{\vect{R}})}^2 } \bar\proptime } \right]_{s=-1/2} \\ 
    &  - \frac{(\hbar \mu)^2 R_{cl}^3}{2 u_0} \int \dd{\bar{\vect{R}}} \abs{\Sigma_0(\bar{\vect{R}})}^2.
  \end{aligned}
\end{equation}
The experimentally tunable $s$-wave scattering length mentioned at the beginning is simply given by $a_s = \frac{m u}{2 \pi \hbar^2}$, and it diverges at the Feshbach resonance, indicating that the system has reached its critical (or unitary, in this context) point~\cite{Strinati:2018wdg}.
In our scheme it is not necessary to renormalize the coupling of \(\Sigma_0^2\),
and criticality simply corresponds to the limit \(u_0 \to 0\).%

We want to express the result in the form of an integral over space, which is by construction well-defined only for \( 1- \bar V(\bar{\vect{R}}) < \delta\).
At leading order, however, the result is well-defined up to \(R_{\text{cl}}\) and we do not need any extra terms, which would otherwise lead to an unphysical scheme dependence.

\newcommand*{\hypF}{\prescript{}{2}{F}_{1}}
If the integral over \(\bar\proptime\) is performed first, we obtain integrals of the form
\begin{equation}\label{eq:polesThatNeedReg}
I_{m,n}(y) = - \eval*{ \frac{\Gamma(n-1/2)}{4 \sqrt{2}\pi^{7/2}} \int_0^{\infty} \dd{q} q^{2+m} \pqty*{\pqty*{q^2 - 1}^2 + y^2}^{1/2-n}}_{\text{reg}},
\end{equation}
which are well-defined for \(n > 1/2 \) and \(-3 < m < 4n -5\), but can be analytically continued to other values in terms of Gaussian hypergeometric functions \(\hypF\):
\begin{multline}
  I_{m,n}(y) = - \frac{y^{5/2 + m/2 - 2 n}}{2 (2 \pi)^{7/2}} \Bigg[ \Gamma(-\tfrac{3+m}{4}) \Gamma(n - \tfrac{5+m}{4})\hypF(- \tfrac{m+1}{4}, n - \tfrac{m+5}{4}, \tfrac{1}{2},-\tfrac{1}{y^2 })  \\
  + \frac{2}{y}  \Gamma(-\tfrac{5+m}{4}) \Gamma(n - \tfrac{3+m}{4})\hypF(\tfrac{m-1}{4}, n - \tfrac{m+3}{4}, \tfrac{3}{2},-\tfrac{1}{y^2 }) \Bigg] .
\end{multline}

At leading order we find
\begin{equation} \label{eq:OmegaLO}
\begin{aligned}
    \Omega_{LO}(\mu)
    & = -4 \pi \hbar \mu R_{cl}^3 \left( \frac{m \mu}{\hbar} \right)^{\frac{3}{2}}  \int \dd{\bar{\vect{R}}} I_{0,0}\pqty*{\tfrac{\Sigma_{0}(\bar{\vect{R}})}{1- V(\bar{\vect{R}})} } \left( 1 - \bar V(\bar{\vect{R}}) \right)^{\frac{5}{2}} - \frac{(\hbar \mu)^2 R_{cl}^3}{2 u_0} \int \dd{\bar{\vect{R}}} \abs{\Sigma_0(\bar{\vect{R}})}^2\\
    & = -4 \pi \frac{m^{\frac{3}{2}}}{\hbar^3}  \int \dd{\vect{R}} I_{0,0}\pqty*{\tfrac{\Sigma_{0}(\bar{\vect{R}})}{1- V(\bar{\vect{R}})} } \left( \hbar \mu - V(\vect{R}) \right)^{\frac{5}{2}} - \frac{(\hbar \mu)^2 R_{cl}^3}{2 u_0} \int \dd{\bar{\vect{R}}} \abs{\Sigma_0(\bar{\vect{R}})}^2.
\end{aligned}
\end{equation}
where
\begin{equation}
  I_{0,0}(y) = - \frac{y^{5/2}}{2 (2 \pi)^{7/2}} \bqty*{ \Gamma(- \tfrac{5}{4}) \Gamma(\tfrac{3}{4}) \hypF(- \tfrac{5}{4}, -\tfrac{1}{4}, \tfrac{1}{2}; - \tfrac{1}{y^2}) + \frac{2}{y} \Gamma(- \tfrac{3}{4}) \Gamma(\tfrac{5}{4}) \hypF(- \tfrac{3}{4}, \tfrac{1}{4}, \tfrac{3}{2}; - \tfrac{1}{y^2})   } . 
\end{equation}

\begin{figure}
  \centering
  \includegraphics[width=.5\textwidth]{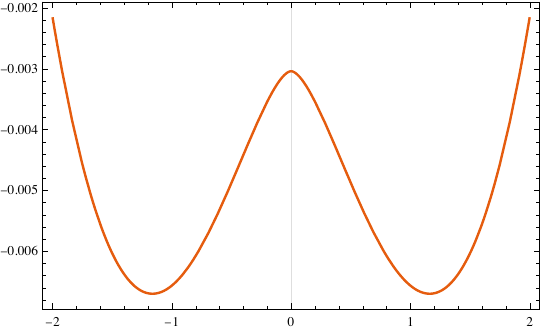}
  \caption{Spontaneous symmetry breaking. The effective action evaluated at the saddle has a minimum for a non-zero value of the parameter \(y\) that controls the expectation value of the gap. Shown is $-I_{0,0}(y)$, which appears in the effective action.}
  \label{fig:SSB}
\end{figure}

The saddle equation at this order is
\begin{equation}
  \label{eq:leading-saddle}
    \Sigma_{0}(\bar{\vect{R}}) I_{0,1}\pqty*{\tfrac{\Sigma_{0}(\bar{\vect{R}})}{1- V(\bar{\vect{R}})}} = 0 ,
\end{equation}
where we have used the recursion equation
\begin{equation}
  \frac{d}{dy}I_{m,n}(y) = -2 y I_{m,n+1}(y) ,
\end{equation}
and admits the solution
\begin{equation}
    \Sigma_{0}(\bar{\vect{R}}) = y_0 \pqty*{1- V(\bar{\vect{R}})},
\end{equation}
with \(y_0 \approx 1.1622\dots \) (see also Figure~\ref{fig:SSB}).
This analysis confirms the predicted form of $\Sigma_0(\bar{\vect{R}})$, Eq.~\eqref{eq:Sigma0_LO}.
At large \(N\) and for a large chemical potential $\mu$, there is an explicit realization of the Cooper mechanism (see Eq.~\eqref{eq:StrataDef} and below): the Stratonovich field \(\sigma(\tau, \vect{r})\) acquires a non-trivial \ac{vev} which spontaneously breaks the $U(1)$ symmetry. The system is then in a superfluid phase, which is the underlying assumption of the large-charge \ac{eft}, allowing an explicit computation of the Wilsonian coefficients.

At the saddle we obtain
\begin{equation}
  c_0 = 4\pi\, I_{0,0}(y_0)  \approx 0.0841\dots,
\end{equation}
which, using Eq.~\eqref{eq:C0_VS_Bertsch}, allows us to extract the value of the Bertsch parameter:
\begin{equation}\label{eq:number-Bertsch}
    \xi \approx 0.5906\dots .
\end{equation}
In the absence of an external potential, this value was obtained via mean-field theory in the celebrated work~\cite{Eagles:1969zz}, and agrees with the large-$N$ computations in~\cite{Veillette:2007zz}, in which the \ac{nlo}-in-$N$ correction was also computed.

Next, recall that in Eq.~\eqref{eq:OmegaLO}, $R_{cl}^3 (2 m \mu / h)^\frac{3}{2} = \varepsilon^{-3} \gg 1$ is the large-charge regime. In fact, the leading dependence of the charge on the chemical potential (at criticality) is given by
\begin{equation}
    Q_{LO} = -\frac{1}{\hbar} \frac{\del \Omega_{LO}^\text{crit.}(\mu)}{\del \mu}
    = \varepsilon^{-3} \cdot \frac{5 c_0}{2} \int \dd{\bar{\vect{R}}} \left( 1 - \bar V(\bar{\vect{R}}) \right)^\frac{3}{2} \gg 1,
\end{equation}
which justifies a posteriori this terminology. Specializing to the case of the harmonic potential $V(\vect{r}) = \frac{m \omega^2}{2} \vect{r}^2$ (\emph{i.e.}, $\bar V(\bar{\vect{R}}) = \bar{\vect{R}}^2$) with $\omega \ll \mu$, we find
\begin{equation}
    Q_{LO} = \left( \frac{\mu}{\omega} \right)^3 \frac{5 \pi^2 c_0}{2^\frac{5}{2}} = \left( \frac{\mu}{\omega} \right)^3 \frac{1}{3 \xi^\frac{3}{2}} = \left( \frac{\mu}{\omega} \right)^3 \cdot 0.734\ldots
\end{equation}
Using Eq.~\eqref{eq:DeltaQ_vs_FreeEnergy}, the leading dependence of the conformal dimension of the large-charge operator on the charge $Q$ is therefore given by
\begin{equation}
    \Delta_{LO}(Q) = \frac{3}{4} \left[ \frac{5 \pi^2 c_0}{2^\frac{5}{2}} \right]^{-3} Q^\frac{4}{3} = \frac{3^\frac{4}{3}}{4} \sqrt{\xi} Q^\frac{4}{3} = Q^\frac{4}{3} \cdot 1.893\ldots
\end{equation}

\subsection{Subleading corrections}%

By inspection of Eq.~\eqref{eq:HeatK_Problem_OrderN}, it is easy to show that the components of the heat kernel are of the form
\begin{equation}
\begin{dcases}
    \Kdiag_n(\bar{\vect{R}}, \bar{\vect{p}}; \bar\proptime) & \equiv e^{-\bbdiag_0(\bar{\vect{R}}, \bar{\vect{p}}) \bar\proptime} \sum_{l=0}^{n+1} \bar\proptime^l \Kdiag_{n,l}(\bar{\vect{R}}, \bar{\vect{p}}), \\
    \Koff_n(\bar{\vect{R}}, \bar{\vect{p}}; \bar\proptime) & \equiv e^{-\bbdiag_0(\bar{\vect{R}}, \bar{\vect{p}}) \bar\proptime} \sum_{l=0}^{n+1} \bar\proptime^l \Koff_{n,l}(\bar{\vect{R}}, \bar{\vect{p}}),
\end{dcases}
\end{equation}
where $\Kdiag_{n,0}(\bar{\vect{R}}, \bar{\vect{p}}) = \delta_{0n}$ and $\Koff_{n,0}(\bar{\vect{R}}, \bar{\vect{p}}) = 0$. That is, these components are polynomials in $\bar\proptime$ of degree (at most) $n+1$, multiplied by an overall factor of $e^{-\bbdiag_0 \bar\proptime}$.

With this, the trace in Eq.~\eqref{eq:S0_DimLessStuff} can be decomposed explicitly order by order in \(\varepsilon\), and the integrals over momentum and proper time can be performed and expressed in terms of the functions \(I_{m,n}\).
At criticality,
\begin{equation}
\begin{aligned}
    \frac{\Omega_{N^nLO}^\text{crit.}(\mu)}{\hbar \mu R_{cl}^3 (2 m \mu / \hbar)^\frac{3}{2}}
     = -\sum_{l=0}^{n+1} \int \dd{\bar{\vect{R}}} \left[ \int \frac{\dd{\bar\proptime}}{\Gamma(s) \bar\proptime} \bar\proptime^{s+l} \int \frac{ \dd{\bar{\vect{p}}}}{(2 \pi)^3} \Kdiag_{n,l}(\bar{\vect{R}}, \bar{\vect{p}}) \cdot e^{-\bbdiag_0(\bar{\vect{R}}, \bar{\vect{p}}) \bar\proptime} \right]_{s=-\frac{1}{2}} ,
\end{aligned}
\end{equation}
which is subject to the gap equation,
\begin{equation}
  \frac{\delta \Omega_{N^nLO}^\text{crit.}}{\delta \Sigma_0} = 0.
\end{equation}
Once the above saddle equation is accounted for, this recipe gives the order-$\varepsilon^n$ \emph{bulk} contribution to the zero-temperature limit of the grand-canonical potential at criticality (and one still needs to account for the edge contributions).
\subsection{Next-to-leading order}%
\label{sec:next-to-leading-order-bulk}

At next order in \(\varepsilon\), \emph{i.e.}, $n = 1$ in Eq.~\eqref{eq:HeatK_Problem_OrderN}, we use the fact that $\poisson*{\bbdiag_0, \Kdiag_0} = 0$, so that the heat kernel equations at this order can be put into the form
\begin{equation}
    \begin{cases}
        \del_{\bar\proptime} \hat \Kdiag_1 + \bbdiag_1 = 0, \\
        \del_{\bar\proptime} \hat \Koff_1 + \bboff_1 = 0,
    \end{cases}
\end{equation}
with
\begin{equation}
	\begin{cases}
		\hat \Kdiag_1(\bar{\vect{R}},\bar{\vect{p}}; 0) = 0, \\
		\hat \Koff_1(\bar{\vect{R}},\bar{\vect{p}}; 0) = 0,
	\end{cases}
\end{equation}
where the hat notation means
\begin{align}
    \Kdiag_i & \equiv \Kdiag_0 \cdot \hat \Kdiag_i, &  \Koff_i & \equiv \Kdiag_0 \cdot \hat \Koff_i.
\end{align}
Thus, the first correction to the leading-order heat kernel is
\begin{equation}
    K_1(\bar{\vect{R}}, \bar{\vect{p}}; \bar\proptime) = -\bar\proptime
    e^{-\bbdiag_0(\bar{\vect{R}}, \bar{\vect{p}}) \bar\proptime}
    \begin{pmatrix}
    \Sigma_{0}(\bar{\vect{R}})^* \Sigma_{1}(\bar{\vect{R}}) + \Sigma_{1}(\bar{\vect{R}})^* \Sigma_{0}(\bar{\vect{R}})
    & 2 i \bar{\vect{p}} \nabla_{\bar{\vect{R}}} \Sigma_{0}(\bar{\vect{R}})
    \\
    -2 i \bar{\vect{p}} \nabla_{\bar{\vect{R}}} \Sigma_{0}(\bar{\vect{R}})^*
    & \Sigma_{0}(\bar{\vect{R}})^* \Sigma_{1}(\bar{\vect{R}}) + \Sigma_{1}(\bar{\vect{R}})^* \Sigma_{0}(\bar{\vect{R}})
    \end{pmatrix},
\end{equation}
and we have
 $\bbdiag_1 = \Sigma_0^* \Sigma_1 + \Sigma_1^* \Sigma_0$, so we obtain
\begin{equation}
  \frac{\Omega_{NLO}^\text{crit.}(\mu)}{\hbar \mu R_{cl}^3 \left( 2 m \mu / \hbar \right)^\frac{3}{2}} = - 4 \pi \int \dd{\bar{\vect{R}}} \left( \Sigma_0(\bar{\vect{R}})^* \Sigma_1(\bar{\vect{R}}) + \Sigma_1(\bar{\vect{R}})^* \Sigma_0(\bar{\vect{R}}) \right) I_{0,1}\pqty*{\tfrac{\abs{\Sigma_{0}(\bar{\vect{R}})}}{1- V(\bar{\vect{R}})}}.
\end{equation}
The gap equation is
\begin{equation}
  \Sigma_{1}(\bar{\vect{R}}) \pqty*{I_{0,1}\pqty*{\tfrac{\abs{\Sigma_{0}(\bar{\vect{R}})}}{1- V(\bar{\vect{R}})}} - 2 \frac{ \Sigma_{0}(\bar{\vect{R}})^2 }{1- V(\bar{\vect{R}})}  I_{0,2}\pqty*{\tfrac{\abs{\Sigma_{0}(\bar{\vect{R}})}}{1- V(\bar{\vect{R}})} } } = -2 \Sigma_{1}(\bar{\vect{R}}) \pqty*{1- V(\bar{\vect{R}})}  y_0^2 I_{0,2}(y_0) = 0,
\end{equation}
which admits the single solution $\Sigma_1 = 0$.
We find that the first corrections in \(\varepsilon\) to the gap and to the grand potential both vanish, in agreement with the \ac{eft} prediction:
\begin{equation}
  \Omega_{NLO}(\mu) = 0 .
\end{equation}

\subsection{Next-to-next-to-leading order}%
\label{sec:next-to-next-to-leading-order-bulk}

At second order in \(\varepsilon\), or \ac{nnlo}, the heat kernel conditions become
\begin{equation}
  \begin{cases}
    \del_{\bar\proptime} \hat \Kdiag_2 + \bbdiag_2 - \pqty*{ \bbdiag_1^2 + \bboff_1^2 + \ppoisson{\bbdiag_0, \bbdiag_0} } \bar\proptime + \frac{1}{8} \Pi_{IJ} \Pi_{LM} \del_{IL} \bbdiag_0 \del_J \bbdiag_0 \del_M \bbdiag_0 \bar\proptime^2 = 0, \\
    \del_{\bar\proptime} \hat \Koff_2 + \bboff_2 \bar\proptime + \bbdiag_1 \bboff_1 \bar\proptime^2 = 0,
  \end{cases}
\end{equation}
and  the \ac{nnlo} action takes the form
\begin{equation}
\begin{aligned}
 \frac{\Omega_{NNLO}^\text{crit.}(\mu)}{\hbar \mu R_{cl}^3 \left( 2 m \mu / \hbar \right)^\frac{3}{2}} = {}& \int \dd{\bar{\vect{R}}} \frac{4 \pi}{9 \pqty*{1-\bar V}^{5/2}} \biggl( 4 (5 I_{2,3}-3 I_{0,3}) (\Sigma_0)^2 (\nabla \Sigma_0)^2 \\
  &\kern-50pt +\pqty*{1-\bar V}^3 ((9 I_{0,1}-18 I_{0,2}+48 I_{2,2}-8 I_{2,3}-30 I_{4,2}+8 (3 I_{4,3}-3 I_{6,3}+I_{8,3})) \Laplacian \bar V\\
  &\kern-50pt -18 I_{0,1} ((\Sigma_1)^2+2 \Sigma_0 \Sigma_2))+2 \pqty*{1-\bar V}^2 ((9 I_{0,2}-9 I_{2,2}+4 (I_{2,3}-2 I_{4,3}+I_{6,3})) (\nabla \Sigma_0)^2 \\
  &\kern-50pt +(9 I_{0,2}-15 I_{2,2}+4 (I_{2,3}-2 I_{4,3}+I_{6,3})) \Sigma_0 \Laplacian \Sigma_0\\
  &\kern-50pt +3 (3 I_{0,2}-2 I_{0,3}-3 I_{2,2}+6 I_{2,3}-6 I_{4,3}+2 I_{6,3}) (\nabla \bar V)^2) \\
  &\kern-50pt +12 \pqty*{1-\bar V} \Sigma_0 (3 I_{0,2} \Sigma_0 (\Sigma_1)^2+2 (I_{0,3}-2 I_{2,3}+I_{4,3}) \nabla \bar V \nabla  \Sigma_0)\biggr) .
  \end{aligned}
\end{equation}
As expected, the coefficient of \(\Sigma_2\) is proportional to  \(\Sigma_0 I_{0,1}\), the leading-order variation in Eq.~\eqref{eq:leading-saddle}, and vanishes on shell.

The gap equation is obtained by varying with respect to \(\Sigma_0\):
\begin{equation}
  \frac{\delta}{\delta \Sigma_0} \Omega = \bqty*{\frac{\del}{\del \Sigma_0} - \nabla \frac{\del}{\del \nabla \Sigma_0} + \Laplacian \frac{\del}{\del \Laplacian  \Sigma_0} } \Omega = 0,
\end{equation}
and, using the relations between the \(I_{m,n}\) ,
\begin{equation}
	I_{m,n}(y) = (n-1/2) \pqty*{I_{m+4,n+1} -2 I_{m+2, n+1} + (1+y^2) I_{m,n+1}},
\end{equation}
we find the profile
\begin{equation} \label{eq:StrataNNLO}
  \sigma_0(\tau, \vect{r}) = y_0 \pqty{\hbar\mu - V(\vect{r})} + y_1 \frac{\hbar^2}{m} \frac{(\nabla V(\vect{r}))^2}{\pqty{\hbar\mu - V(\vect{r})}^2} + y_2 \frac{\hbar^2}{m} \frac{\Laplacian{V(\vect{r})}}{\hbar\mu - V(\vect{r})} + \dots,
\end{equation}
with 
\begin{equation}
    y_1\approx -0.004347\dots, \qquad y_2\approx -0.1608\dots.
\end{equation}
This is the first non-trivial correction to the profile of the Stratonovich field.
We stress once more that this expression is only valid up to a distance \(\delta\) from the edge at \(\bar{\vect{R}} = 1\).
We will discuss in the next section what happens closer to the edge of the cloud, where we will need to merge our small-gradient expansion with the perturbation theory of the boundary dynamics, subject to appropriate matching conditions.

As for the effective action, we find\footnote{Notably, we can express the action in terms of rational
coefficients of $I_{0,0}(y_0)$.}
\begin{equation}
\begin{aligned}
  S =  \frac{4\pi}{\hbar^3}\beta m^{3/2} I_{0,0}(y_0) \int \dd{\vect{r}} \biggl( 
  &\pqty*{\hbar\mu - V(\vect{r})}^{5/2} + \frac{5}{64} \frac{\hbar^2}{m} \frac{(\nabla V)^2}{\sqrt{\hbar\mu - V(\vect{r})}} \\
  & - \frac{25}{48} \frac{\hbar^2}{m}  \Laplacian V(\vect{r}) \sqrt{\hbar\mu - V(\vect{r})} 
  \biggr).
\end{aligned}
\end{equation}
From this, we can read off the Wilsonian parameters:
\begin{align}%
  \label{eq:result-c1c2}
  c_1 &= \frac{5 \pi}{16}I_{0,0}(y_0) \approx 0.006577\dots, & c_2 &= \frac{25 \pi}{108} I_{0,0}(y_0) \approx 0.004872\dots .
\end{align}
This gives the final result for the scaling dimension of the lowest operator of charge $Q$
\begin{equation}%
  \label{eq:resultDelta}
  \frac{\Delta}{N} =  0.8313 \pqty*{\frac{Q}{N} }^{4/3} + 0.2631 \pqty*{\frac{Q}{N} }^{2/3} + \dots
\end{equation}
The value for $c_2$ satisfies the bound of~\cite{Son:2005rv} on the transverse response function:  The transverse response itself must be negative for a stable condensate, requiring that $c_2 > 0$.  The fact that $2c_1 + 3c_2 > 0$ also places us in the regime were, in the zero-temperature, low-momentum limit, in the absence of an external potential or boundary dynamics, $1 \to 2$ phonon splitting is forbidden.  %

It is important to stress that the form that we find for the gap and the Lagrangian density is precisely the one expected based on locality and dimensional arguments: as long as we only look at the bulk, the problem has two scales, $\mu$ and $\nabla V$, so any physical quantity of dimension \(\Delta(G)\), in the limit of large $\mu$, must take the form
\begin{equation}
	G(\mu, \vect r) = \pqty*{ \hbar \mu-V(\vect r)}^{\Delta(G)/2} \mathscr{F}\pqty*{\tfrac{\hbar}{m^{1/2}}\tfrac{ \nabla V(\vect r)}{(\hbar \mu-V(\vect r))^{3/2}}}. 
\end{equation}

\section{The edge expansion}%
\label{sec:Edge}

In the previous section we obtained an asymptotic expansion of the bulk contributions to the free energy in terms of the small parameter $\varepsilon \sim \omega/\mu$.
These contributions match precisely the predictions from the bulk \ac{eft}, and allow us to compute the corresponding Wilsonian coefficients at leading order in $1/N$. 
We have noted, however, that this expansion is only valid in the bulk, up to some distance $\delta$ away from the point where the particle density vanishes at leading order in $\varepsilon$.
This is the typical setup studied in boundary-layer theory~\cite{bender1999advanced}. The domain of a differential equation is split into two (or more) regions. In each region, one finds an asymptotic expansion of the solution, which may then be matched over an intermediate region where both approximations are valid.
In the case where exact solutions are known across multiple regions, one is instructed to \emph{patch} these solutions smoothly at transition points.  Crucially, this is to be distinguished from cases where only asymptotic expansions are known in different regions, wherein one must instead perform a \emph{matching}.
That is, given two such solutions supported in different regions, one needs to find an overlapping region (the \emph{intermediate limit}) in which the two asymptotic expansions have the same functional form.

Being able to find such a matching imposes a constraint on the thickness $\delta$ of the boundary layer. Qualitatively speaking, the boundary layer has to be sufficiently thick that the corresponding asymptotic solution (controlled by $\delta$) is rich enough to permit a matching with the functional form coming from the outer (bulk) solution. Concretely, $\delta$ is fixed by a dominant-balance argument in which two or more of the terms of the one-particle Hamiltonian must be of the same order, allowing us to retain enough information to perform a matching. This identifies the so-called \emph{distinguished limit}.
For simplicity, here we will consider the case of a spherically-symmetric potential.
Correspondingly, the boundary layer is a spherical shell around $\bar r \equiv \abs{\bar{\vect{r}}} = 1$, so we introduce the \emph{inner} variable $\bar u$ as
\begin{equation}
  \bar{r} = 1 - \bar u \delta.
\end{equation}

In the limit of \(\delta \to 0\), the metric is approximately flat in terms of a radial and two orthogonal directions.
To see that, start from \(\setR^3\) in polar coordinates,
\begin{equation}
  \dd{s}^2 = \dd{\bar{r}}^2 + \bar{r}^2 \pqty*{ \frac{\dd{z}^2}{1- z^2} + \pqty{1 - z^2} \dd{\phi}^2} ,
\end{equation}
with \(z \in (-1,1)\) and \(\phi \in (0, 2 \pi)\). Rescaling \(z = \bar x \delta\) and \(\phi = \bar y \delta\), so that \(\bar x \in(- 1/\delta, 1/\delta)\) and \(\bar y \in (0, 2 \pi/\delta)\), the metric becomes
\begin{equation}
  \dd{s}^2 = \delta^2 \pqty*{ \dd{\bar u} + \dd{\bar x}^2 + \dd{\bar y}^2} = \delta^2 \pqty*{\dd{\bar u} + \dd{\bar x_{\perp}}^2}  + \order{\delta}
\end{equation}
and the Laplacian is approximately
\begin{equation}
  \Laplacian_{\bar r}{} = \frac{1}{\delta^2} \pqty*{\Laplacian_{\perp}{} + \frac{\dd^{2}{}}{\dd{u^{2}}}} + \order{1/\delta}.
\end{equation}

Since $1 - \bar V(1) = 0$ by the definition of $R_{cl}$, the dimensionless Hamiltonian $\bar h(\bar r)$ becomes
\begin{equation}
	\eval*{\bar h(\bar r)}_{\bar r =1 - \bar u \delta} = -\frac{\varepsilon^2}{\delta^2}\pqty*{ \Laplacian_{\perp}{} + \frac{\dd^{2}{}}{\dd{u^{2}}}} - \bar u \delta \del_{\bar r} \bar V(1) + \order{(\bar u \delta)^2}.
\end{equation}
The only possible dominant balance is obtained when the first two terms are of the same order, \emph{i.e.},
\begin{equation}
	\frac{\varepsilon^2}{\delta^3} = \order{1}.
\end{equation}
The double-scaling limit $\varepsilon \to 0$, $\delta \to 0$, such that $\varepsilon^2/\delta^3=1$, is our \emph{distinguished limit} in which the operator $\bar h$ is
\begin{equation}
  \eval*{\bar h(\bar r)}_{\bar r =1 - \bar u \delta} = -\delta \cdot \pqty*{ \Laplacian_{\perp}{} + \frac{\dd^{2}{}}{\dd{u^{2}}} + \alpha \bar u } + \order{(\bar u \delta)^2},
\end{equation}
where \(\alpha = \bar V'(1)\) and, in particular, for the harmonic oscillator \(\alpha = 2\).
This is the Airy operator, corresponding to the linearization of the confining potential at the turning point.

From the point of view of the gap profile, in contrast to the bulk condition in Eq.~\eqref{eq:bulk-condition}, the edge condition is that \(\sigma\) varies on the same scale as the one fixed by its own \ac{vev}
\begin{align}
   \sigma(\vect{r}) \lessapprox \frac{\hbar^2}{2 m}\frac{\pqty*{\del_{\vect{r}} \sigma(\vect{r})}^2}{\sigma(\vect{r})^2}  && \text{(edge condition).}
\end{align}
In the edge region%
\footnote{In the usual boundary theory language, the edge is the \emph{inner region} and the bulk is the \emph{outer region}.}%
, then, we expect to be able to write the Stratonovich field as an expansion in \(\delta\)
\begin{equation}
	\eval*{ \bar\sigma(\bar u)}_{\text{edge}} = \bar\sigma_{\text{edge}}^{(1)}(\bar u) \delta + \bar\sigma_{\text{edge}}^{(2)}(\bar u) \delta^2  + \dots,
\end{equation}
so that the condition becomes
\begin{equation}
    \bar \sigma_{\text{edge}}^{{1}}(\bar u) \lessapprox \frac{\epsilon^2}{\delta^3} \frac{\pqty*{\del_{\bar u}  \bar \sigma_{\text{edge}}^{{1}}(\bar u)}^2}{ \bar \sigma_{\text{edge}}^{{1}}(\bar u)^2},
\end{equation}
which is satisfied if \(\bar \sigma_{\text{edge}}^{{1}}(\bar u)\) and its \(\bar u\) derivative are of order \(\order{1}\).

The boundary condition for \(\bar \sigma_{\text{edge}}\) is obtained by imposing that in a region where both the bulk and edge approximations are consistent, the known result from the bulk is matched.
Let us start from the bulk expansion for the Stratonovich field, Eq.~\eqref{eq:StrataNNLO}.
In the spherically-symmetric case we have found that
\begin{equation}
  \eval*{\bar\sigma(\bar{r})}_{\text{bulk}} = y_0 \pqty*{1 - \bar V(\bar{r})} + \frac{2}{\varepsilon^2} \left[ y_1 \frac{\bar V'(\bar{r})^2}{\pqty*{1 - \bar V(\bar{r})}^2} + y_2 \frac{\bar V''(\bar{r})}{1 - \bar V(\bar{r})} \right] + \dots
\end{equation}
In the \emph{intermediate limit} $\delta \to 0$, $\bar u \to \infty$ with $ \bar u \delta \to 0$, for the harmonic oscillator potential, this becomes
\begin{equation}
    \eval*{ \bar \sigma(1 - \bar u \delta)}_{\text{bulk}} = \delta \cdot \pqty*{2 y_0 \bar u +\frac{y_1}{\bar u^2} + \dots } + \order{\delta^2},
\end{equation}
which also scales like $\delta$ at leading order. To be able to perform the matching, the edge solution $\bar\sigma_{\text{edge}}$ must have the same functional form for $\delta \to 0$, and we must have
\begin{equation}
	\sigma_{\text{edge}}^{(1)}(\bar u) \underset{\bar u \to \infty}{\sim} 2 y_0 \bar u +\frac{y_1}{\bar u^2} + \dots .
\end{equation}

We are now able to write the expression of the \ac{bdg} operator close to the boundary:
\begin{equation}
  B = \hbar \mu \delta \bar B^{(1)}(\bar u) + \order{\mu \delta^2} = 
    \hbar \mu \delta \begin{pmatrix}
         \Laplacian_{\perp}{} + \frac{\dd^{2}{}}{\dd{u^{2}}} + 2 \bar u & \sigma_{\text{edge}}^{(1)}(\bar u) \\
      \sigma_{\text{edge}}^{(1)}(\bar u)^* &  - \Laplacian_{\perp}{} - \frac{\dd^{2}{}}{\dd{u^{2}}} - 2 \bar u \end{pmatrix} + \dots .
\end{equation}
To compute the edge contribution to the free energy we need to solve the gap equation
\begin{equation}
  \label{eq:8}
  \begin{dcases}
    \frac{\delta \Tr*{\abs{B}}}{\delta \sigma_{\text{edge}}^{(1)}(\bar u)} = 0 \\
    \sigma_{\text{edge}}^{(1)}(\bar u) \underset{\bar u \to \infty}{\sim} 2 y_0 \bar u +\frac{y_1}{\bar u^2} + \dots \\
    \sigma_{\text{edge}}^{(1)}(\bar u) \xrightarrow[\bar u \to -\infty]{}0 ,
  \end{dcases}
\end{equation}
to obtain the saddle profile \(\sigma_{\text{edge}}^{(1)}(\bar u) = \braket{\sigma_{\text{edge}}^{(1)}(\bar u)}\), and then evaluate the corresponding value of the trace of the operator \(\abs{\bar B^{(1)}}\), 
\begin{equation}
  F_{\text{edge}} = \frac{\hbar \mu \delta }{2} \Tr*{\abs{\bar B^{(1)}(\bar u)}}_{\braket{\sigma_{\text{edge}}^{(1)}(u)}} + \dots ,
\end{equation}
where the trace is understood over the edge region.
This problem is more challenging than the bulk calculation discussed in the previous section, since here we have no small parameter by which to control a perturbative expansion, and a deeper
analysis is beyond the scope of the present work.
Rather, we will limit ourselves to estimating the order in \(\mu\) and \(\omega\) at which this contribution enters the expression for the free energy, to confirm the prediction of the \ac{eft} in~\cite{Hellerman:2020eff}.
The first observation is that the trace is divergent and requires regulation.
At first sight, the Airy linear potential is concerning because it seems to imply a continuous spectrum.
However, we are ultimately computing the trace of the absolute value \(\abs{\bar B^{(1)}} = \pqty{(B^{(1)})^2}^{1/2}\), and \((B^{(1)})^2\) is bounded from below and increases without bound in the \(u\) direction, so that its spectrum is manifestly discrete.
Thus, we can use a standard zeta-function regulator and write the free energy as the Mellin transform of the heat kernel at coincident points
\begin{equation}
  \Tr*{\abs{\bar B^{(1)}(\bar u)}} = \eval*{\frac{1}{ \Gamma(s)} \int \frac{\dd{\bar\proptime}}{\bar\proptime} \bar\proptime^s \Tr*[e^{- \abs{\bar B^{(1)}} \bar\proptime}] }_{s=-1} ,
\end{equation}
which is well-defined.
Inverting the order of integration, the free energy is written as the integral of a local density,
\begin{equation}
  \bar b^{(1)}(\bar x, \bar y, \bar u) = \eval*{\frac{1}{\Gamma(s)}\int \frac{\dd{\bar\proptime}}{\bar\proptime} \bar\proptime^s \braket{\bar x, \bar y, \bar u|e^{- \abs{\bar B^{(1)} } \bar\proptime}|\bar x, \bar y, \bar u} }_{s=-1} ,
\end{equation}
which is parametrically of order \(\order{1}\), because it does not depend on \(\mu\) or \(\omega\).
This allows us to compute the parametric dependence of \(F\):
\begin{equation}
  \begin{aligned}
    F_{\text{edge}} &= \frac{\hbar \mu \delta }{2} \Tr*{\abs{\bar B^{(1)}}} = \frac{\hbar \mu \delta }{2} \int_{-1/\delta}^{1/\delta} \dd{\bar x} \int_0^{2\pi/\delta} \dd{\bar y} \int_0^1 \dd{\bar u} \bar b^{(1)}(\bar x, \bar y, \bar u) 
    \\
                    &= \hbar \order*{\frac{\mu}{\delta}} = \hbar \order*{\mu^{4/3} R_{\text{cl}}^{2/3}},
  \end{aligned}
\end{equation}
where we have used the fact that the integral over \(\bar b^{(1)}\) is of order \(1/\delta^2\).
In the case of the harmonic trap we find
\begin{equation}
  \eval*{ F_{\text{edge}}}_{\text{ht}} = \frac{\hbar^{4/3}}{m^{1/3}} \order*{ \frac{\mu^{5/3}}{\omega^{2/3}} } + \dots ,
\end{equation}
which is precisely the expected scaling of the leading-order edge contribution from the \ac{eft}~\cite{Hellerman:2020eff}.

\section{Conclusions}%
\label{sec:Conclusions}

In this work, we have studied the unitary Fermi gas in a trapping potential in $3+1$ dimensions at zero temperature in the limit of large charge and large number of Fermion flavors $N$.
This is a well-justified approximation at the microscopic level, providing a foundation for the Thomas--Fermi approximation at leading order. We present a clean algorithmic procedure for extracting data from perturbation theory, requiring no other inputs beyond dimensional and symmetry arguments and basic techniques of quantum mechanics. 

Previous attempts have involved a variety of approximation techniques, which we put here on a sound theoretical basis:
The large N limit turns the one-loop calculation into a justified approximation at \(N = \infty\) (which is typically extrapolated to \(N=1\) in mean field theory); the large-charge limit provides us with a small parameter proportional to the ratio of the coupling \(\omega\) of the harmonic potential and the chemical potential \(\epsilon = \omega/(2\mu)\), which allows us to perform a Moyal (gradient) expansion of the functional determinant resulting from integrating out the original microscopic \ac{dof}.
In this way we have found a consistent microscopic derivation that confirms the \ac{eft} prediction for the form of the bulk operators, their contribution to the energy, and the form of the gap equation.
We have also derived the first three Wilsonian coefficients at leading order in N and \ac{nnlo} in $\omega/\mu$.
Though the bulk expansion breaks down near the droplet edge, we were able to estimate the parametric dependenceof the leading-order edge contributions to the energy starting from first principles, confirming the \ac{eft} prediction of~\cite{Hellerman:2020eff}.
We reserve the precise determination of the numerical coefficient for future work.

\medskip
Further characterizing the Wilsonian coefficients in the boundary theory constitutes a clear area for further investigation.
Beyond this, one can also imagine expanding the reach of our theoretical understanding of the broader system by relaxing the limits of the regime studied herein:  For instance, by including $1/N$ corrections, or by leaving the unitary point along the lines of~\cite{Orlando:2021usz,Moser:2021bes}, which would provide another controlled setting for the analysis discussed in~\cite{Chowdhury:2023ahp}, etc.
Perhaps most importantly, it would be important to connect with experimental observations. As a first step, one might consider the doubly-integrated axial density~\cite{PhysRevLett.92.120401}, which can be computed based on the results herein.
On the other hand, it would be valuable to obtain experimental data (of the density profile, perhaps) with sufficient resolution to probe the concrete predictions set forth here, particularly in the regime beyond the reach of the Thomas--Fermi approximation.

\medskip

  \section*{Acknowledgments}

  \begin{small}
  The authors would like to thank Nathan Lundblad and Martin Zwierlein for useful discussions
  pertaining to experimental realizations of the unitary Fermi gas.
    The work of S.H. is supported by
    the World Premier International Research Center Initiative (WPI Initiative), MEXT, Japan;
    by the JSPS Program for Advancing Strategic International Networks to Accelerate the
    Circulation of Talented Researchers; and also supported in part by JSPJ KAKENHI Grant
    Numbers JP22740153, JP26400242.%
    The work of V.P. and S.R. is supported by the Swiss National Science Foundation under grant number 200021\_192137.
    S.H. and V.P. also thank the
Simons Center for Geometry and Physics for hospitality while this work
was in progress.
  \end{small}

\appendix

\section{An alternative regularization}

As noted in the main text in Eq.~\eqref{eq:polesThatNeedReg}, the functions \(I_{m,n}(y)\) need to be analytically continued beyond \(n > 1/2\), \(-3 < m < 4n - 5\).
In particular, the function appearing in the leading-order term, \(I_{0,0}(y)\), is divergent.
The regularization typically used in the \ac{bcs} literature goes as follows. Setting \(n = m = 0\), one finds
\begin{equation} \label{eq:OmegaLO_a_la_BCS}
    \Omega_{LO}(\mu) = - \hbar \mu R_{cl}^3 \int \dd{\bar{\vect{R}}} \bqty*{ \frac{\hbar \mu}{2 u_0} \abs{\Sigma_0(\bar{\vect{R}})}^2 
     + \left( \frac{2 m \mu}{\hbar} \right)^\frac{3}{2} \int \frac{\dd{\bar{\vect{p}}}}{(2 \pi)^3} \sqrt{ \pqty*{1 - \bar V(\bar{\vect{R}}) - \bar{\vect{p}}^2}^2 + \abs{\Sigma_0(\bar{\vect{R}})}^2 } },
\end{equation}
where we have replaced $\varepsilon$ by its explicit expression, via Eq.~\eqref{eq:EpsilonFixed}.
  For large values of \(\bar{\vect{p}}\), the integrand diverges as
\begin{equation}
    \bar{\vect{p}}^2 \sqrt{\pqty*{1 - \bar V(\bar{\vect{R}}) - \bar{\vect{p}}^2}^2 + \abs{\Sigma_0(\bar{\vect{R}})}^2 } = -\bar{\vect{p}}^2 \pqty*{ 1 - \bar V(\bar{\vect{R}}) - \bar{\vect{p}}^2} + \frac{1}{2} \abs{\Sigma_0(\bar{\vect{R}})}^2 + \order*{1/\bar{\vect{p}}^2}.
\end{equation}
The first term corresponds to the integrand evaluated at $\Sigma_0 = 0$ and can be removed normalizing the partition function by that of the free theory.
The second term combines with the bare coupling $u_0$, to define the renormalized coupling as
\begin{equation}
    \frac{1}{u} \equiv \frac{1}{u_0} + \left( \frac{2 m \mu}{\hbar} \right)^{\frac{3}{2}} \frac{1}{\hbar \mu} \int \frac{\dd{\bar{\vect{p}}}}{(2 \pi)^3} \frac{1}{\bar{\vect{p}}^2} = \frac{1}{u_0} + \int \frac{\dd{\vect{p}}}{(2 \pi \hbar)^3} \frac{2 m}{\vect{p}^2},
\end{equation}
where, in the second equality, we restored the dimensionful momentum $\vect{p} \equiv (2 m \hbar \mu)^{\frac{1}{2}} \bar{\vect{p}}$ for a more direct comparison with the standard result (see for example~\cite{Veillette:2007zz,Strinati:2018wdg}).

\setstretch{1}

\printbibliography{}

\end{document}